\documentclass[12pt, draftclsnofoot, onecolumn]{IEEEtran}
\usepackage{epsfig}
\usepackage{amssymb}
\usepackage{bm}
\usepackage[font=footnotesize]{caption} % small
\usepackage{latexsym,graphicx}
\usepackage{tabulary}
\usepackage{longtable}
\usepackage{algpseudocode} 
\usepackage{algorithm}
\usepackage{array}
\usepackage{mathrsfs}
\usepackage{color}
\definecolor{myc1}{rgb}{0,0,1}
\usepackage{amsmath,graphicx}
\usepackage{relsize}
\usepackage{epsfig}
\usepackage{setspace}
\usepackage{cite}
\usepackage{mathtools}

\usepackage{hyperref}
\usepackage{amssymb}
\usepackage{tabularx}
\usepackage{caption}
\usepackage{subcaption}
\usepackage[normalem]{ulem}
\usepackage{soul,xcolor}
 \usepackage{nopageno}
\def\BibTeX{{\rm B\kern-.05em{\sc i\kern-.025em b}\kern-.08em
    T\kern-.1667em\lower.7ex\hbox{E}\kern-.125emX}}

 \makeatletter
 \let\NAT@parse\undefined
 \makeatother

\begin{document}

%\title{Square Deviation Based Symbol-Level Selective Relaying for Full-Duplex Cooperative Network} 
%\title{Asymptotic Performance Analysis of Joint User Association and FD-UAV Deployment in Spectrum Shared Networks}
\title{\huge Joint Beamforming, User Association, and Height Control for Cellular-Enabled UAV Communications}
\author{Jiancao Hou, \IEEEmembership{Member, IEEE,} Yansha Deng, \IEEEmembership{Member, IEEE,} and\\Mohammad Shikh-Bahaei, \IEEEmembership{Senior Member, IEEE}% <-this % stops a space
%\thanks{This work was partly funded by the ICT-248894 WHERE 2.}
% <-this % stops a space
%\thanks{This work was supported by the Engineering and Physical Science Research Council (EPSRC) through the SENSE grant EP/P003486/1.

%J. Hou and M. Shikh-Bahaei are with the Department of Informatics, King's College London, London, United Kingdom, WC2B 4BG. E-mail: \{jiancao.hou, m.sbahaei\}@kcl.ac.uk. The material in this paper has been in part submitted to XX Conference XX year \cite{Hou2019}.}
}
\markboth{} {Shell \MakeLowercase{\textit{et al.}}: Bare Demo of
IEEEtran.cls for Journals}\maketitle

\begin{abstract} 
%Summary of your paper: brief introduction, motivation, and contributions.
Supporting reliable and seamless mobility for aerial users, such as unmanned aerial vehicles (UAVs), is a key challenge for the next-generation  cellular systems. To tackle this challenge, we propose a joint beamforming, user association, and UAV-height control framework for cellular-connected multi-UAV networks with multiple antenna base stations (BSs).  With the aim of maximizing the minimum achievable rate for UAVs subject to  co-existed terrestrial user equipment's rate constraints, we devise a hierarchical bi-layer iterative algorithm to optimize BSs' beamforming vectors, UAV association matrix, and the height of UAVs jointly. With the aid of projection gradient method in inner layer iteration and geometric program modelling plus convex-concave procedure in outer layer iteration, our proposed algorithm is proved to converge to a local optimum. Taking mobility characteristics of UAVs into account, we also exploit our proposed algorithm for imperfect channel estimation scenario. Numerical results show that our proposed algorithm can achieve improved UAVs' minimum achievable rate compared with that of the conventional nearest association of UAVs for both perfect and imperfect channel estimation scenarios. Moreover, we also examine the trade-off between the UAVs' minimum achievable rate and the frequency for updating optimization variables with single moving UAV.
\end{abstract}


\IEEEpeerreviewmaketitle
%%%%%%%%%%%%%%%%%%%%%%%%%%%%%%%%%%%%%%%%%%%%%%%%%%%%%%%%%%%%%%%%%%%%%%
 
                         %%%I. Introduction%%% 

%%%%%%%%%%%%%%%%%%%%%%%%%%%%%%%%%%%%%%%%%%%%%%%%%%%%%%%%%%%%%%%%%%%%%%
\section{Introduction}
Due to the ability of flexible on-demand deployment and high line-of-sight (LoS) probability, unmanned aerial vehicles (UAVs) have become appealing solutions, which will likely open attractive vertical markets for a wide range of applications, such as aerial inspection/rescue, cargo delivery, surveillance, and precision agriculture, etc\cite{Zeng2016,Zeng2019,Mozaffari2018,Amorim2017}. However, traditional UAV communications mainly rely on simple and direct connections between the UAV and the dedicated ground user via unlicensed frequency band, where data rate, security, and operational coverage can hardly be guaranteed. To overcome these limitations and provide reliable and seamless connectivity, UAVs need to be controlled and connected over a wider area wireless network, where terrestrial cellular network is well positioned to provide the services due to its pervasive deployment and guaranteed accessibility \cite{Amorim2017,3GPP2017}. %facilitate on-demand deployment, rapid expansion of applications and services, new generations of wireless communications are expected to rely on even higher dynamic resource allocation, wider range of coverage, and ultra-reliable and much lower transmission latency\cite{Saad2019,Zeng2016,Towhidlou2018}. Among the various promising techniques,

Despite the potential advantages might be brought via cellular-connected UAVs, the integration of UAVs to the traditional terrestrial networks are also envisioned to bring important challenges to the considered systems. Specifically, with ability of flying in a three-dimension geographic area for UAVs, their favourable LoS propagation conditions could be one of their strongest limiting factors, since the significant LoS interference generated by UAVs via uplink transmission can deteriorate the co-existed ground user equipment's (G-UE) rates either in the same cell or neighbouring cell. Conversely, the downlink transmission of ground base station (G-BS) can also generate severe interference to the UAV associated with the neighbouring G-BS \cite{Rodriguez2019,3GPP2017}. Note that, contemporary G-BSs in cellular network are designed to support reliable connectivity for G-UEs with the nearest user association process \cite{Geraci2018,Stanczak2018,Yajnanarayana2018}. However, when one or many UAVs present and tend to associate to G-BSs, due to strong co-channel interference and dynamically changes of fading channels, the nearest user association may not lead to the optimal system performances. In this case, how to handle the co-channel interference and user association with respect to UAV mobility robustness will bring challenges \cite{3GPP2017,Qualcomm2017,Huawei2017,Lin2018}.

In light of these challenges, the authors in \cite{Azari2017} studied the feasibility of integrating UAVs into existing cellular networks by presenting a generic framework for evaluating the coverage performance. The results demonstrated that the favourable propagation condition that UAVs enjoy due to their altitudes is also their strongest limiting factor, and the negative effect can be substantially reduced by optimizing UAV flying altitude, G-BS height and antenna down-tilted angle. Meanwhile, the authors in \cite{Zhang2019} took UAV mobility into account with the target to minimize the UAV's mission completion time by optimizing its trajectory by the means of graph theory. The mobility challenges, such as handover's reliability/latency and mobility performance, for cellular-enabled UAV communications was evaluated in \cite{Stanczak2018} via simulations, where the results revealed that handover/radio link failure can be directly linked with the height of UAV, and advanced interference cancellation techniques should be complemented with additional enhancements to improve mobility robustness. It is noted that the above solutions for providing a fast and reliable connection between UAV and cellular network only considered single-antenna based G-BSs, where spatial multiplexing and/or diversity gains have not been exploited. 
 % In \cite{Amer2019}, the authors shown that the coverage probability of cooperative cellular-connected UAVs network can be improved by increasing the altitude of UAVs.
 
Consider UAVs and G-UEs co-exist within the same time and frequency, and G-BSs are equipped with multiple antennas. The authors in \cite{Geraci2018} evaluated the performances of cellular-connected downlink UAVs communications supported by massive multiple input multiple output (MIMO) enabled network with zero-forcing (ZF) beamforming, where users tend to associate to a nearby G-BS. In \cite{Amer2019}, the authors derived the successful content delivery probability of a cellular-enabled UAV network and shown that exploiting conjugate beamforming from massive MIMO enabled G-BSs to spatially multiplex an UAV and G-UEs can substantially improve the UAV's performance. Such scheme assumed perfect channel state information (CSI) at G-BSs. Note that, both \cite{Geraci2018} and \cite{Amer2019} assumed that only massive MIMO combining with simple beamforming methods was exploited. In practice, due to the imperfect channel estimation and the size of communication equipments, massive MIMO may not always be feasible solutions. Motivated by above discussion, in this paper, we propose a joint beamforming, user association, and UAV-height control framework to maximize the minimum achievable rate for UAV in cellular-enabled multiple UAVs uplink communication systems with limited number of antennas per G-BS. Moreover, in consideration of mobility of UAVs, we further study the proposed algorithm by taking imperfect CSI into account.

The main contributions of this paper can be summarized as follows:
\begin{itemize} 
\item  We formulate the minimum achievable rate for UAV optimisation problem with G-UE's target rate constraints being guaranteed via jointly optimizing the UAV association matrix, G-BSs' beamforming vectors, and the height of the UAVs. To solve this mixed integer non-linear optimization problem, a hierarchical bi-layer searching algorithm is proposed to find the optimal solution iteratively. Based on this, we also analyze the computational complexity and convergence of the proposed algorithm theoretically.
\item The proposed algorithm consists of two main iterations: In outer layer iteration, we fix the height of UAVs and jointly optimize UAV association matrix and beamforming vectors with helps of bi-section search and projection gradient method. Then, given the UAV association matrix, we jointly optimize the height of UAVs and beamforming vectors by exploiting geometric program modeling and convex-concave procedure in inner layer iteration. 
\item Taking into account mobility of UAVs, we relax the perfect CSI assumption and study the proposed scheme in statistical channel environments, i.e., each G-BS has an estimated version plus the estimation error covariance of its related channel links. In this case, we derive \textit{effective} SINRs for our imperfect CSI scenario. Numerical results show that the proposed algorithm outperforms the conventional UAV association via the nearest G-BS in terms of the minimum achievable rate for UAV, especially when co-channel interference effects is small. In addition, considering one of UAVs is moving, the proposed algorithm can provide much stable rate performance than the conventional UAV association policy.  
\end{itemize}

The rest of the paper is organized as follows. Section II presents system model and problem formulation of the proposed UAV-enabled MIMO cellular network. Section III provides the proposed hierarchical bi-layer search method to solve the formulated objective accompanied with its computational complexity and convergence analysis. Then, we extend the proposed scheme working in imperfect CSI condition in Section IV. Section V provides the numerical results and the corresponding discussion. Section VI concludes the paper. Throughout the paper, the main notations are summarized in Tab.~I.

\begin{table}[t]
\center
{\footnotesize
\caption{Summarizes the basic notations in the paper}\label{tab1}
\begin{tabular}{ |p{1.2cm}||p{8.5cm}|}
%\hline
%\multicolumn{2}{|c|}{Notation List} \\
\hline
\textbf{Symbol} & \textbf{Usage} \\
\hline
$\mathbb{R}^{N}$, $\mathbb{C}^{N}$ & The set of real and complex $N$-tuples, respectively.\\
\hline
$\mathbf{a}$, $\mathbf{A}$ &  The column vector and matrix, respectively. \\
\hline
$[\mathbf{A}]_{m,n}$ & The element in the $m^{\mathrm{th}}$ row and the $n^{\mathrm{th}}$ column of matrix $\mathbf{A}$. \\
\hline
$\mathcal{A}\cup\mathcal{B}$ & The union of set $\mathcal{A}$ and set $\mathcal{B}$.\\
\hline
$|\cdot|$ & The absolute value of a scalar.\\
\hline
$\|\cdot\|$ & Euclidean norm of a vector or Frobenius norm of a matrix.\\
\hline
$\|\cdot\|_{1}$ & The taxicab norm of a vector.\\
\hline
$\mathbb{E}\{\cdot\}$ & The expectation of random variable(s).\\
\hline
$(\cdot)^{-1}$ & Inverse of a matrix. \\
\hline
$\min\{a,b\}$  & The minimum value between $a$ and $b$. \\
\hline
$\{\cdot$\} & A set. \\
\hline
$\mathbf{I}_{N}$  & The identity matrix with size of $N\times N$.\\
\hline 
$\mathbf{1}_{N}$  & The all one column vector with size of $N\times1$.\\
\hline
$(\cdot)^{H}$ & The vector (or matrix) conjugate transpose.  \\
\hline
$(\cdot)^{T}$ & The vector transpose.  \\
\hline
\end{tabular}}
\end{table}
%Here and throughout, lower case boldface symbols are used to denote column vectors (e.g. $\mathbf{a}$), while upper case boldface symbols are used to denote matrices (e.g. $\mathbf{A}$). $(\cdot)^{H}$ denotes vector (or matrix) conjugate transpose. $\mathcal{C}^{N}$ denotes the set of complex $N$-tuples. $\ceil{x}$ and $\floor{x}$ denote the smallest integer greater than or equal to $x$ and the largest integer less than or equal to $x$, respectively. $\mathrm{rank}(\cdot)$ denotes the rank of a matrix. $|\cdot|$ denotes an absolute value. $\|\cdot\|$ denotes Euclidean norm of a vector or Frobenius norm of a matrix. $\mathbb{E}\{\cdot\}$ denotes expectation of random variable(s). $(\cdot)^{-1}$ denotes inverse of a matrix, and $\mathrm{diag}(\mathbf{a})$ denotes a diagonal matrix whose diagonal is vector $\mathbf{a}$.
%%%%%%%%%%%%%%%%%%%%%%%%%%%%%%%%%%%%%%%%%%%%%%%%%%%%%%%%%%%%%%%%%%%%%%

                     %%%II. System Model%%% 

%%%%%%%%%%%%%%%%%%%%%%%%%%%%%%%%%%%%%%%%%%%%%%%%%%%%%%%%%%%%%%%%%%%%%%
\section{System Model and Problem Formulation}
% Provide a comprehensive description of the system model which leads to your theoretical analysis. A block diagram is usually demanded to clarify your % presentation. Formulate the problem you want to resolve. Please note this system model should be much related to your simulations.
\begin{figure}[t]
\begin{center}
\epsfig{figure=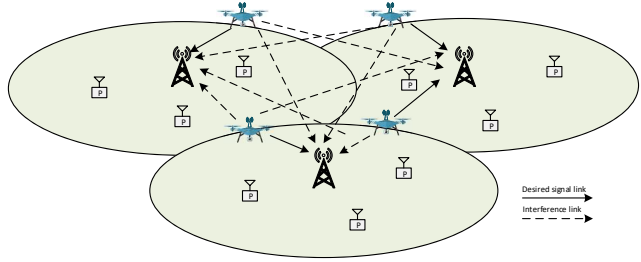,scale=0.55,angle=0}
\end{center}
\caption{An example of multi-UAV co-existed cellular network.}\label{F1a}
\end{figure}
As shown in Fig.~\ref{F1a}, we consider the uplink transmission of cellular-connected UAV networks, where multiple UAVs co-exist with the G-UEs to upload data to their associated G-BSs. The $G$ G-BSs are assumed with equal height of $h_{\mathrm{G}}$, with its set denoted as $\mathcal{G}=\left\{1,\ldots,G\right\}$. Each G-BS is equipped with $N$ antennas and serves $K$ pre-associated single-antenna G-UEs. Without loss of generality, we assume the set of G-UEs as $\mathcal{K}=\{\underbrace{1,\ldots,K}_{\mathcal{K}_1},\ldots,\underbrace{(G-1)K+1,\ldots,GK}_{\mathcal{K}_{G}}\}$  in the network, where $\mathcal{K}_{i}$ is a group of G-UEs associated with the $i^{\mathrm{th}}$ G-BS. We assume that there are $U$ single-antenna UAVs flying above the ground, where each UAV can only associate with one G-BS, and the set of UAVs is denoted as $\mathcal{U}=\left\{1,\ldots,U\right\}$. By exploiting the spatial multiplexing gain, each G-BS with $N$ receive antennas is allowed to associate with a maximum of $N-K$ UAVs as long as its associated $K$ G-UEs' target rate constraints are not violated. Let's denote $\mathcal{U}_{i}(\subset\mathcal{U})$ as the set of UAVs associated with the $i^{\mathrm{th}}$ G-BS, where $U_{i}$ is defined as the cardinality of the set $\mathcal{U}_{i}$. Combining with $K$ pre-associated G-UEs per cell, the total number of served UEs by the $i^{\mathrm{th}}$ G-BS can be expressed as $J_i\triangleq K+U_{i}$, where $J_{i}\leq N$ and its set is denoted as $\mathcal{J}_{i}=\mathcal{K}_{i}\cup\mathcal{U}_{i}$. It is assumed that the transmissions from UAVs and G-UEs to G-BSs are with universal time and frequency reuse, and the G-BSs estimate their associated channel links including the location of UAVs by exploiting the pilot-based channel estimation method within fixed coherent time \cite{Bjornson2016}.

\subsection{Channel Models}
The channel links between G-UEs and G-BSs follow standard territorial cellular channel models, which consist of the non Line-of-Sight (NLoS) large-scale fading and the small-scale multi-path fading. To avoid collisions and follow air traffic regulations, we assume the flying heights of UAVs are no less than 100 meters and no more than 300 meters. In this case, by complying with channel models defined in 3GPP Release 15 \cite{3GPP2017}, the channel links between UAVs and G-BSs are characterized as LoS large-scale fading and small-scale fading. In general, the large scale fading between the $i^{\mathrm{th}}$ G-BS and the $j^{\mathrm{th}}$ UE can be expressed as
\begin{IEEEeqnarray}{ll}\label{SecII-A-01}
\zeta_{v}(r_{i,j})=A_{v}d_{i,j}^{-\alpha_{v}}=A_{v}\left[r^2_{i,j}+(h_{j}-h_{\mathrm{G}})^2\right]^{-\frac{\alpha_{v}}{2}},~i\in\mathcal{G},~j\in\mathcal{K}\cup\mathcal{U},
\end{IEEEeqnarray}
where $A_{v}$, $v\in\{\mathrm{L},\mathrm{N}\}$, is a constant representing the path loss at the reference distance $d_{i,j}=1$ for either LoS or NLoS link. $r_{i,j}$ is the horizontal distance between the $i^{\mathrm{th}}$ G-BS and the $j^{\mathrm{th}}$ UE, and $h_{j}$ is the vertical height of the $j^{\mathrm{th}}$ UE. Here, if $j\in\mathcal{K}$, then these UEs are G-UEs, such that $h_{j}=0$; if $j\in\mathcal{U}$, then these UEs are UAV, such that $h_{\mathrm{max}}=300\text{m}\geq h_{k}\geq h_{\mathrm{min}}=100$m. $\alpha_{v}$, $v\in\{\mathrm{L},\mathrm{N}\}$, is the path loss exponents for either LoS or NLoS link. The small scale fading channels are formulated by Nakagami-$m$ model, e.g., $\mathbf{h}_{i,j}\in\mathbb{C}^{M}$, where the channel gain, e.g., $\Omega_{v}$, $v\in\{\mathrm{L},\mathrm{N}\}$, for each entry follows a Gamma distribution with probability density function (p.d.f.) as
\begin{equation}\label{SecII-A-02}
f_{\Omega_v}(\omega)=\frac{m^{m_{v}}_{v}\omega^{m_{v}-1}}{\Gamma(m_{v})}\mathrm{exp}(-m_v\omega).
\end{equation}
In \eqref{SecII-A-02}, $m_{v}$, $v\in\{\mathrm{L},\mathrm{N}\}$, is the fading parameters for either LoS or NLoS link, and it is assumed to be integers for analytical tractability. 
%Since the height of UAVs decides whether the channel is prominent with LoS, the probability of LoS of the $i^{\mathrm{th}}$ G-BS at horizontal distance $r_{i,k}$ from the $k^{\mathrm{th}}$ user is given by [ref]
%\begin{equation}\label{eq03}
%\mathcal{P}_{\mathrm{LoS}}(r_{i,k})=\prod^{m}_{n=0}\left[1-\mathrm{exp}\left(-\frac{\left[h_\mathrm{G}-\frac{(n+0.5)(h_\mathrm{G}-h_\mathrm{U})}{m+1}\right]^{2}}{2c^2}\right)\right],~i\in\mathcal{G},~k\in\mathcal{U},
%\end{equation}
%where $m=\floor{\frac{r_{i,k}\sqrt{ab}}{1000}-1}$; $a$, $b$ and $c$ are the pre-set parameters based on the environments. Here, $\mathcal{P}_{\mathrm{LoS}}(r_{i,k})$ is a decreasing function of $r_{i,k}$ and an increasing function of $h_\mathrm{U}$, and we have $\mathcal{P}_{\mathrm{NLoS}}(r_{i,k})=1-\mathcal{P}_{\mathrm{LoS}}(r_{i,k})$. Moreover, by following the work in [Al-Hourani2014WCL], the trend of LoS probability can be closely approximated to a simple modified Sigmoid function (S-curve) of the following form:
%\begin{equation}\label{eq04}
%\mathcal{P}_{\mathrm{LoS}}(\theta_{i,jk})=\frac{1}{1+x\mathrm{exp}(-y[\theta-x])},
%\end{equation}
%where $x$ and $y$ are called here the S-curve parameters. In addition, we have $r_{i,k}=(h_{\mathrm{U}}-h_{\mathrm{G}})/\mathrm{tan(\theta_{\mathit{i,k}})}$. Such approximation significantly ease the calculation of the LoS probability, and also it allows the theoretical analysis more feasible.

%%%%%%%%%%%%%%%%%%%%%%%%%%%%%%%%%%%
\subsection{Power Control and User Association}
Following the standard, the G-UEs and UAVs perform uplink transmission using the statistics-aware uplink power control as in \cite{Bjornson2016,3GPP2017}. Such power control takes the average over the received power, which is easier for implementation, especially for high mobility UAVs communication systems. With given location information of the G-UEs, the transmit power of the $k^{\mathrm{th}}$ G-UE served by the $i^{\mathrm{th}}$ G-BS can be formulated as
\begin{IEEEeqnarray}{ll}\label{SecII-B-01}
P_{k}=\min\{\overline{P}_{\mathrm{K}},B_{\mathrm{N}}\cdot(r^{2}_{i,k}+h_{\mathrm{G}}^2)^{\frac{\alpha_{\mathrm{N}}}{2}}\},~i\in\mathcal{G},~k\in\mathcal{K}_{i},
\end{IEEEeqnarray}
where $B_{\mathrm{N}}>0$ is a cell-specific parameter for G-UE with NLoS, which is used to limit the transmit power no more than its amplifiers to handle. $\overline{P}_{\mathrm{K}}$ is the maximum transmit power constraint of a G-UE; $r_{i,k}$ is the horizontal distance between the $k^{\mathrm{th}}$ G-UE and its associated G-BS. For the $u^{\mathrm{th}}$ UAV associated with the $i^{\mathrm{th}}$ G-BS, its transmit power can be formulated as
\begin{equation}\label{SecII-B-02}
P_{i,u}=\min\{\overline{P}_{\mathrm{U}},B_{\mathrm{L}}\cdot\left[r^{2}_{i,u}+(h_{u}-h_{\mathrm{G}})^2\right]^{\frac{\alpha_{\mathrm{L}}}{2}}\},~i\in\mathcal{G},~u\in\mathcal{U}_{i},
\end{equation}
where, similar to $B_{\mathrm{N}}$ in \eqref{SecII-B-01}, $B_{\mathrm{L}}>0$ is a cell-specific parameter for UAV with LoS. $\overline{P}_{\mathrm{U}}$ is the maximum transmit power constraint of a UAV. $r_{i,u}$ is the horizontal distance between the $u^{\mathrm{th}}$ UAV and its associated G-BS. Let's define an association matrix $\mathbf{A}$ with the size of $G\times U$, where its elements (i.e., $a_{g,u}\in\{0,1\}$) represent the association status between the $u^{\mathrm{th}}\in[1,\ldots, U]$ UAV and the $g^{\mathrm{th}}\in[1,\ldots,G]$ G-BS. Here, $a_{g,u}=1$ means the $u^{\mathrm{th}}$ UAV is associated with the $g^{\mathrm{th}}$ G-BS, otherwise $a_{g,u}=0$. The received signal $\mathbf{y}_{i}\in\mathbb{C}^{M}$ at the $i^{\mathrm{th}}$ G-BS can be modeled as
\begin{equation}\label{SecII-B-03}
\mathbf{y}_{i}= \sum_{k\in\mathcal{K}}\sqrt{P_{k}\zeta_{\mathrm{N}}(r_{i,k})}\mathbf{h}_{i,k}s_{k}+\sum_{g\in\mathcal{G}}\sum_{u\in\mathcal{U}}a_{g,u}\sqrt{P_{g,u}\zeta_{\mathrm{L}}(r_{i,u})}\mathbf{h}_{i,u}s_{u}+\mathbf{v}_{i},
\end{equation}
where $s_{k}\in\mathbb{C}$ and $s_{u}\in\mathbb{C}$ are the symbols transmitted by the $k^{\mathrm{th}}$ G-UE and the $u^{\mathrm{th}}$ UAV with unit power, respectively, with $\mathbb{E}\{|s_{k}|^2\}=1,\forall k\in\mathcal{K}$ and $\mathbb{E}\{|s_{u}|^2\}=1,\forall u\in\mathcal{U}$. In addition, $\mathbf{v}_{i}\in\mathbb{C}^{M}$ is the additive white Gaussian noise (AWGN) with zero mean and variance $\sigma^{2}_{0}$. In \eqref{SecII-B-03}, the first summation term represents the signals received from G-UEs, and the second summation term is the received signal from UAVs. 

Introducing the $i^{\mathrm{th}}$ G-BS generated beamforming vectors $\mathbf{z}_{j}\in\mathbb{C}^{M}$, $\forall j\in\mathcal{J}_{i}$, $\forall i\in\mathcal{G}$, for co-channel interference mitigation and/or desired signal enhancement, the decoded signal at the $i^{\mathrm{th}}$ G-BS from the $j^{\mathrm{th}}$ its associated UE is given by
\begin{equation}\label{SecII-B-04}
\mathbf{z}^{\mathrm{H}}_{j}\mathbf{y}_{i}=\sum_{k\in\mathcal{K}}\sqrt{P_{k}\zeta_{\mathrm{N}}(r_{i,k})}\mathbf{z}^{\mathrm{H}}_{j}\mathbf{h}_{i,k}s_{k}+\sum_{g\in\mathcal{G}}\sum_{u\in\mathcal{U}}a_{g,u}\sqrt{P_{g,u}\zeta_{\mathrm{L}}(r_{i,u})}\mathbf{z}^{\mathrm{H}}_{j}\mathbf{h}_{i,u}s_{u}+\mathbf{z}^{\mathrm{H}}_{j}\mathbf{v}_{i},
\end{equation}
where $j$ could be a G-UE or a UAV that is associated with the $i^{\mathrm{th}}$ G-BS. For simplicity, we denote $\mathcal{D}^{(j)}_{i,k}\triangleq P_{k}\zeta_{\mathrm{N}}(r_{i,k})|\mathbf{z}^{H}_{j}\mathbf{h}_{i,k}|^2$, $\mathcal{F}^{(j)}_{i,u}\triangleq\zeta_{\mathrm{L}}(r_{i,u})|\mathbf{z}^{H}_{j}\mathbf{h}_{i,u}|^2$, and $\mathcal{N}^{(j)}_{i}\triangleq|\mathbf{z}^{\mathrm{H}}_{j}\mathbf{v}_{i}|^2$. The received SINR at the $i^{\mathrm{th}}$ G-BS from its served $k^{\mathrm{th}}$ G-UE can be written as
\begin{equation}\label{SecII-B-05}
\mathrm{SINR}_{i,k}= \frac{\mathcal{D}^{(k)}_{i,k}}{\sum_{k^{'}\in\mathcal{K}/k}\mathcal{D}^{(k)}_{i,k^{'}}+\sum_{g\in\mathcal{G}}\sum_{u\in\mathcal{U}}a_{g,u}P_{g,u}\mathcal{F}^{(k)}_{i,u}+\mathcal{N}^{(k)}_{i}}.
\end{equation}
Moreover, the received SINR at the $i^{\mathrm{th}}$ G-BS from the $u^{\mathrm{th}}$ UAV can be written as
\begin{equation}\label{SecII-B-06}
\mathrm{SINR}_{i,u}= \frac{a_{i,u}P_{i,u}\mathcal{F}^{(u)}_{i,u}}{\sum_{k\in\mathcal{K}}\mathcal{D}^{(u)}_{i,k}+\sum_{g\in\mathcal{G}}\sum_{u^{'}\in\mathcal{U}/u}a_{g,u^{'}}P_{g,u^{'}}\mathcal{F}^{(u)}_{i,u^{'}}+\mathcal{N}^{(u)}_{i}}.
\end{equation}
In \eqref{SecII-B-05} and \eqref{SecII-B-06}, the first summation term in the denominator consists of both intra- and inter- cells interference from G-UEs, and the second summation term in the denominator consists of the interference from UAVs. Note that, a UAV formulates its transmit power according to the distance to its associated G-BS. Different heights of the UAV and/or G-BS associations give different power allocation and then generate different interference power to other UEs. %Such procedure will further complicate the problem in comparison to the ones in literature.  
%%%%%%%%%%%%%%%%%%%%%%%%%%%%%%%%%%%%%%%
\subsection{Problem Formulation}
Following the above system descriptions, our main aim in this paper is to answer the following question: for given target rate of individual G-UE in a cell, what will be the optimal strategy for the UAV association, MIMO receive beamforming design, and UAVs height setup, which can maximize the minimum achievable rate of the UAVs? Mathematically, we can formulate the objective problem as 
\begin{IEEEeqnarray}{ll}\nonumber
\mathcal{P}1:~\underset{\mathbf{A},\{h_{u}\}_{u\in\mathcal{U}},\{\mathbf{z}_{j}\}_{j\in\mathcal{K}\cup\mathcal{U}}}{\mathrm{max}}&~t\\
~~~~~~~~~~~~~~~~~\mathrm{s.t.}&~\mathrm{SINR}_{i,u}\geq a_{i,u}t,~\forall i\in\mathcal{G},~\forall u\in\mathcal{U},
\IEEEyesnumber
\IEEEyessubnumber\label{SecII-C-01a}\\
&~\mathrm{SINR}_{i,k}\geq \gamma,~\forall i\in\mathcal{G}, \forall k\in\mathcal{K}_{i},
\IEEEyesnumber
\IEEEyessubnumber\label{SecII-C-01b}\\
&~\sum^{G}_{i=1}a_{i,u}= 1,~\forall u\in\mathcal{U},
\IEEEyesnumber
\IEEEyessubnumber\label{SecII-C-01c}\\
&~\sum^{U}_{u=1}a_{i,u}\leq N-K,~\forall i\in\mathcal{G},
\IEEEyesnumber
\IEEEyessubnumber\label{SecII-C-01d}\\
&~h_{\mathrm{min}}\leq h_{u}\leq h_\mathrm{max},~\forall u\in\mathcal{U},
\IEEEyesnumber
\IEEEyessubnumber\label{SecII-C-01e}\\
&~a_{i,u}\in\{0,1\},~\forall i\in\mathcal{G},~\forall u\in\mathcal{U},
\IEEEyesnumber
\IEEEyessubnumber\label{SecII-C-01f}
\end{IEEEeqnarray}
where the first constraint \eqref{SecII-C-01a} in problem $\mathcal{P}1$ is to guarantee UAVs' minimum SINR no less than $t$; the second constraint \eqref{SecII-C-01b} is to guarantee all individual G-UE's target SINR no less than $\gamma$; \eqref{SecII-C-01c} is to limit that each UAV must associate with one and only one G-BS; \eqref{SecII-C-01d} indicates that each G-BS can maximally associate with $N-K$ UAVs due to its limited spatial DoFs; \eqref{SecII-C-01e} is individual UAV's height constraint; \eqref{SecII-C-01f} is the integer value for the association matrix $\mathbf{A}$. It is shown that problem $\mathcal{P}1$ is a mixed integer non-linear optimization problem, which in general is  difficult to handle or find the global optimal solution. This motivate us in the following section to develop the bi-layer search method to solve the problem.% In addition, similar as $P_{i,k}$ in \eqref{SecII-B-01}, $b_{i,k}$ in \eqref{SecII-C-01b} is equal to one if $k\in\left[(i-1)K,\ldots,iK\right]$, otherwise, $b_{i,k}=0$.

%%%%%%%%%%%%%%%%%%%%%%%%%%%%%%%%%%%%%%%%%%%%%%%%%%%%%%%%%%%%%%%%%%%%%%

                     %%%III. System Model%%% 

%%%%%%%%%%%%%%%%%%%%%%%%%%%%%%%%%%%%%%%%%%%%%%%%%%%%%%%%%%%%%%%%%%%%%%
\section{The Proposed Bi-Layer Search Method}
In this section, we mainly focus on solving the problem $\mathcal{P}1$ using our proposed bi-layer search method with the assumption that the perfect CSI of UAVs to G-BSs links and G-UEs to G-BSs links are available at their corresponding G-BSs. Specifically, in outer layer iteration, with a fixed UAVs' heights $h_{u}$, $\forall u\in\mathcal{U}$, we  iteratively optimize the UAV association matrix $\mathbf{A}$ and beamforming vectors $\mathbf{z}_{j}$, $\forall j\in\mathcal{K}\cup\mathcal{U}$, via the proposed bi-section search combining with the gradient projection method \cite{Sanjabi2014}; In inner layer iteration, by giving the UAV association matrix $\mathbf{A}$ from the outer layer iteration, the optimal UAVs' heights $h_{u}$, $\forall u\in\mathcal{U}$, and beamforming vectors $\mathbf{z}_{j}$, $\forall j\in\mathcal{K}\cup\mathcal{U}$, can be jointly obtained with the help of the proposed generalized geometric programming (GGP) \cite{Boyd2007} plus the convex-concave procedure (CCP) \cite{Tao2016}.

%%%%%%%%%%%%%%%%%%%%%%%%%%%%%%%%%%%%%%5
\subsection{Outer Layer Iteration} 
By fixing the UAVs' height $h_{u}$, $\forall u\in\mathcal{U}$, the optimal UAV association matrix $\mathbf{A}$ and beamforming vectors $\mathbf{z}_{j}$, $\forall j\in\mathcal{K}\cup\mathcal{U}$, can be obtained by solving the following sub-problem 
\begin{IEEEeqnarray}{ll}\nonumber
\mathcal{P}1.1:\underset{\mathbf{A},\{\mathbf{z}_{j}\}}{\mathrm{max}}&~t\\
~~~~~~~~~~~\mathrm{s.t.}&~\mathrm{SINR}_{i,u}\geq a_{i,u}t,~\forall i\in\mathcal{G},~\forall u\in\mathcal{U},
\IEEEyesnumber
\IEEEyessubnumber\label{SecIII-A-01a}\\
&~\mathrm{SINR}_{i,k}\geq \gamma,~\forall i\in\mathcal{G},\forall k\in\mathcal{K}_{i},
\IEEEyesnumber
\IEEEyessubnumber\label{SecIII-A-01b}\\
&~\sum^{G}_{i=1}a_{i,u}= 1,~\forall u\in\mathcal{U},
\IEEEyesnumber
\IEEEyessubnumber\label{SecIII-A-01c}\\
&~\sum^{U}_{u=1}a_{i,u}\leq N-K,~\forall i\in\mathcal{G},
\IEEEyesnumber
\IEEEyessubnumber\label{SecIII-A-01d}\\
&~a_{i,u}\in\{0,1\},~\forall i\in\mathcal{G},~\forall u\in\mathcal{U}.
\IEEEyesnumber
\IEEEyessubnumber\label{SecIII-A-01e}
\end{IEEEeqnarray}
Problem $\mathcal{P}1.1$ is a mixed integer non-linear programming problem, which is non-convex and NP-hard due to the coupling among integer elements of association matrix $\mathbf{A}$, beamforming vectors $\mathbf{z}_{j},\forall j$, and the target SINR of UAVs $t$. To solve the problem $\mathcal{P}1.1$, we first need to decouple the $a_{i,u}t$ term on the right-hand side of constraint \eqref{SecIII-A-01a}, where this term represents that the target SINR $t$ of UAV is met only if the $u^{\mathrm{th}}$ UAV is associated with the $i^{\mathrm{th}}$ G-BS. In this case, by introducing the big-$M$ technique \cite{Schrijver1986}, we can rewrite the constraint \eqref{SecIII-A-01a} as
\begin{equation}\label{SecIII-A-02}
\frac{a_{i,u}P_{i,u}\mathcal{F}^{(u)}_{i,u}+M(1-a_{i,u})}{\sum_{k\in\mathcal{K}}\mathcal{D}^{(u)}_{i,k}+\sum_{g\in\mathcal{G}}\sum_{u^{'}\in\mathcal{U}/u}a_{g,u^{'}}P_{g,u^{'}}\mathcal{F}^{(u)}_{i,u^{'}}+\mathcal{N}^{(u)}_{i}}\geq t,
\end{equation}
where $M$ should be a properly large number so that, when $a_{i,u}=0$, constraint \eqref{SecIII-A-01a} is not violated, and when $a_{i,u}=1$, term $M(1-a_{i,u})=0$ makes \eqref{SecIII-A-02} converge back to the original constraint \eqref{SecIII-A-01a}. Such technique aims to find a feasible solution of original problem by introducing an auxiliary variable. The detailed explanation of how to find the proper value of $M$ is presented in Appendix A. It is worth noting that $M$ cannot be arbitrary large, as an arbitrary large $M$ may affect the operation accuracy of computer simulations. 

Following the big-$M$ transformation, the constraint \eqref{SecIII-A-02} is still non-convex due to the coupling among $a_{g,u^{'}}$, $t$ and $\mathbf{z}_{i}$. Moreover, the constraint \eqref{SecIII-A-01b} is also non-convex due to the coupling between $a_{g,u}$ and $\mathbf{z}_{i}$. Apart from those, the problem contains binary variables $a_{i,u}$ (i.e. see \eqref{SecIII-A-01c}), which makes the problem  NP-hard \cite{Garey1990}. Thus, by replacing \eqref{SecIII-A-01a} with \eqref{SecIII-A-02}, we need to implement a sub-layer iterative process within this outer layer iteration to find the optimal $\mathbf{A}$ and $\mathbf{z}_{i},\forall i$ of problem $\mathcal{P}1.1$. Specifically, our proposed sub-layer iterative method is mainly composed of a sub-layer outer iteration and a sub-layer inner iteration, where in this sub-layer outer iteration, the bi-section search is used to find the maximum achievable target SINR $t$ of UAV. Then, in the sub-layer inner iteration, given a particular target $t$, the optimal $\mathbf{A}$ and $\mathbf{z}_{i},\forall i$ can be found by relaxing the binary variables, i.e. $a_{i,u}\in\{0,1\}$ into $a_{i,u}\in[0,1]$, and using a generalized fixed-point method \cite{Yates1995} to determine whether the given $t$ is feasible or not.
%%%%%%%%%%%%%%%%%%%%%%%%%%%%%%%%%
\subsubsection{Sub-Layer Outer Iteration}
Let's use the superscript $(n)$ to denote the iteration number in the outer loop. In addition, we pre-define the lower bound and the upper bound of target SINR of UAV as $\underline{t}$ and $\overline{t}$, respectively. For each iteration, if the result of the previous sub-layer inner iteration loop is `feasible', we will update
\begin{IEEEeqnarray}{ll}\label{SecIII-A-03}
\left\{
\begin{array}{rcr}
t^{(n+1)}&=&\frac{t^{(n)}+\overline{t}}{2},\\
\underline{t}&=&t^{(n)}.\\
\end{array}
\right.
\end{IEEEeqnarray}
Otherwise, if the result of the previous sub-layer inner iteration loop is `infeasible', we have
\begin{IEEEeqnarray}{ll}\label{SecIII-A-04}
\left\{
\begin{array}{rcr}
t^{(n+1)}&=&\frac{t^{(n)}+\underline{t}}{2},\\
\overline{t}&=&t^{(n)}.\\
\end{array}
\right.
\end{IEEEeqnarray}
The outer iteration is terminated when $\overline{t}-\underline{t}\leq\epsilon$, where $\epsilon$ is a small number used for the iteration stop criteria.
%%%%%%%%%%%%%%%%%%%%%%%%%%%%%%%%%%%%%%
\subsubsection{Sub-Layer Inner Iteration}
Given a target SINR $t$ of the UAV from the sub-layer outer iteration, to determine whether it is feasible or not, we shall solve the following problem:
\begin{IEEEeqnarray}{ll}\nonumber
\mathcal{P}1.1.2:~\underset{\mathbf{A},\{\mathbf{z}_{j}\}}{\mathrm{max}}&~\sum^{G}_{i=1}\sum^{U}_{u=1}a_{i,u}\\
~~~~~~~~~~~~~~\mathrm{s.t.}&~\eqref{SecIII-A-02},\eqref{SecIII-A-01b},\eqref{SecIII-A-01d},\eqref{SecIII-A-01e}
\IEEEyesnumber
\IEEEyessubnumber\label{SecIII-A-05a}\\
&~\sum^{G}_{i=1}a_{i,u}\leq 1,~\forall u\in\mathcal{U}.
\IEEEyesnumber
\IEEEyessubnumber\label{SecIII-A-05b}
\end{IEEEeqnarray}
In this problem, we aim to jointly optimize $\mathbf{A}$ and $\{\mathbf{z}_{j}\}_{j\in\mathcal{K}\cup\mathcal{U}}$ in order to find the maximum number of UAVs achieving the minimum target rate $t$. Following our proposed optimization procedure, if each UAV can be associated to a G-BS, i.e., $\sum^{G}_{i=1}\sum^{U}_{u=1}a_{i,u}=U$, the problem $\mathcal{P}1.1.2$ is considered as `feasible', and \eqref{SecIII-A-03} will be used to update $t$ in sub-layer outer iteration. However, if $\sum^{G}_{i=1}\sum^{U}_{u=1}a_{i,u}<U$, the problem $\mathcal{P}1.1.2$ is considered as `infeasible', and \eqref{SecIII-A-04} will be used to update $t$ in sub-layer outer iteration.% It is worth noting that, with a fixed height of UAVs, i.e., $h_{u},\forall u\in{\mathcal{U}}$, problem $\mathcal{P}2$ is equivalent to problem $\mathcal{P}1.1.2$.

Problem $\mathcal{P}1.1.2$ can be solved by firstly relaxing the binary variables of $\mathbf{A}$ to continuous variables, i.e., $a_{i,u}\in[0,1],\forall i,u,$ and then using a generalized fixed-point method, where at each iteration we update either $\mathbf{A}$ or $\{\mathbf{z}_{j}\}_{j\in\mathcal{K}\cup\mathcal{U}}$ while the other variables are kept fixed. More specifically, for a given $\mathbf{A}$, each G-BS first computes 
\begin{IEEEeqnarray}{ll}\label{SecIII-A-06}
\mathbf{z}^{*}_{i,u}&=\arg\underset{\|\mathbf{z}_{i,u}\|=1}{\max}\frac{a_{i,u}B_{\mathrm{L}}A_{\mathrm{L}}\mathbf{z}^{H}_{i,u}\mathbf{h}_{i,u}\mathbf{h}^{H}_{i,u}\mathbf{z}_{i,u}+M(1-a_{i,u})\mathbf{z}^{H}_{i,u}\mathbf{z}_{i,u}}{\mathbf{z}^{H}_{i,u}\left(\sum_{k\in\mathcal{K}}\tilde{\mathcal{D}}_{i,k}+\sum_{g\in\mathcal{G}}\sum_{u^{'}\in\mathcal{U}/u}a_{g,u^{'}}\tilde{P}_{g,u^{'}}\tilde{\mathcal{F}}_{i,u^{'}}+\sigma^{2}_{0}\mathbf{I}\right)\mathbf{z}_{i,u}},
\end{IEEEeqnarray}
and
\begin{IEEEeqnarray}{ll}\label{SecIII-A-07}
\mathbf{z}^{*}_{i,k}&=\arg\underset{\|\mathbf{z}_{i,k}\|=1}{\max}\frac{P_{k}\zeta_{\mathrm{N}}(r_{i,k})\mathbf{z}^{H}_{i,k}\mathbf{h}_{i,k}\mathbf{h}^{H}_{i,k}\mathbf{z}_{i,k}}{\mathbf{z}^{H}_{i,k}\left(\sum_{k^{'}\in\mathcal{K}/k}\tilde{\mathcal{D}}_{i,k^{'}}+\sum_{g\in\mathcal{G}}\sum_{u\in\mathcal{U}}a_{g,u}\tilde{P}_{g,u}\tilde{\mathcal{F}}_{i,u}+\sigma^{2}_{0}\mathbf{I}\right)\mathbf{z}_{i,k}},
\end{IEEEeqnarray}
for all UAVs and G-UEs, where $\tilde{\mathcal{D}}_{i,k}\triangleq P_{k}\zeta_{\mathrm{N}}(r_{i,k})\mathbf{h}_{i,k}\mathbf{h}^{H}_{i,k}$ and $\tilde{\mathcal{F}}_{i,u}\triangleq\zeta_{\mathrm{L}}(r_{i,u^{'}})\mathbf{h}_{i,u}\mathbf{h}^{H}_{i,u}$. In view of the Rayleigh-Ritz quotient result \cite{Golub1996}, the optimal beamforming vector of \eqref{SecIII-A-06} is the eigenvector corresponding to the largest eigenvalue of 
\begin{IEEEeqnarray}{ll}\label{SecIII-A-08}
\left(\sum_{k\in\mathcal{K}}\tilde{\mathcal{D}}_{i,k}+\sum_{g\in\mathcal{G}}\sum_{u^{'}\in\mathcal{U}/u}a_{g,u^{'}}\tilde{P}_{g,u^{'}}\tilde{\mathcal{F}}_{i,u^{'}}+\sigma^{2}_{0}\mathbf{I}\right)^{-1}\cdot\left[a_{i,u}B_{\mathrm{L}}A_{\mathrm{L}}\mathbf{h}_{i,u}\mathbf{h}^{H}_{i,u}+M(1-a_{i,u})\mathbf{I}\right].\nonumber
\end{IEEEeqnarray}
Similarly, the optimal beamforming vector of \eqref{SecIII-A-07} is the eigenvector corresponding to the largest eigenvalue of 
\begin{IEEEeqnarray}{ll}\label{SecIII-A-09}
\left(\sum_{k^{'}\in\mathcal{K}/k}\tilde{\mathcal{D}}_{i,k^{'}}+\sum_{g\in\mathcal{G}}\sum_{u\in\mathcal{U}}a_{g,u}\tilde{P}_{g,u}\tilde{\mathcal{F}}_{i,u}+\sigma^{2}_{0}\mathbf{I}\right)^{-1}\cdot\left[P_{k}\zeta_{\mathrm{N}}(r_{i,k})\mathbf{h}_{i,k}\mathbf{h}^{H}_{i,k}\right].\nonumber
\end{IEEEeqnarray}

Subsequently, the optimal $\mathbf{A}$ with continuous elements can be obtained by solving the relaxed version of the problem $\mathcal{P}1.1.2$ with the fixed $\{\mathbf{z}_{j}\}_{j\in\mathcal{K}\cup\mathcal{U}}$, which becomes a convex problem. It is worth noting that, to optimize the continuous elements of $\mathbf{A}$, we should use the projection gradient descent (PGD) steps instead of directly finding the global optimality of $\mathcal{P}1.1.2$ with fixed $\{\mathbf{z}_{j}\}_{j\in\mathcal{K}\cup\mathcal{U}}$. This is because if we directly find the global optimality at the initial steps, the elements of $\mathbf{A}$ will be fixed during the entire iterative process. In other words, there would be no further optimization of the beamforming vectors $\{\mathbf{z}_{j}\}_{j\in\mathcal{K}\cup\mathcal{U}}$. To elaborate the PGD method, we first reshape the association matrix $\mathbf{A}$ into a vector form (e.g., $\mathbf{a}\in\mathbb{R}^{GU}$) with size of $\mathrm{GU}\times 1$, and define the constraints set in problem $\mathcal{P}1.1.2$ as $\mathcal{Q}$. Starting from a initial point $\mathbf{a}^{(0)}\in\mathcal{Q}$, the PGD iteratively optimizes the following equation until a stopping criteria is met: 
\begin{IEEEeqnarray}{ll}\label{SecIII-A-10} 
\mathbf{a}^{(n+1)} = P_{\mathcal{Q}}\left(\mathbf{a}^{(n)}+\delta\mathbf{1}_{GU}\right),
\end{IEEEeqnarray}
where $\mathbf{1}_{GU}$ is an all one column vector with size of $\mathrm{GU}\times1$. $\delta$ is the carefully chosen step size. $P_{\mathcal{Q}}(\cdot)$ is the projection operator, which is
\begin{IEEEeqnarray}{ll}\label{SecIII-A-11}
P_{\mathcal{Q}}(\mathbf{a}^{(n)})=\arg\underset{\mathbf{a}\in\mathcal{Q}}{\min}\frac{1}{2}\|\mathbf{a}-\mathbf{a}^{(n)}\|^2_{2}.
\end{IEEEeqnarray}
By giving $\mathbf{a}^{(n)}$, the projection process in \eqref{SecIII-A-11} is trying to find a column vector $\mathbf{a}\in\mathcal{Q}$ with the same size as $\mathbf{a}^{(n)}$ which is `closest' to $\mathbf{a}^{(n)}$. After the joint iterative optimization of $\mathbf{A}$ with continuous elements and $\{\mathbf{z}_{j}\}_{j\in\mathcal{K}\cup\mathcal{U}}$ converges, the binary linear programming (BLP) method \cite{Shi1997} can be used to find an optimal solution of the relaxed problem $\mathcal{P}1.1.2$ with the fixed optimized $\{\mathbf{z}_{j}\}_{j\in\mathcal{K}\cup\mathcal{U}}$, for which the assignment variables $\{a_{i,u}\}$ are binary. 

The overall steps to solve problem $\mathcal{P}1.1.2$ is given in \textit{Algorithm 1} in detail. 
\begin{algorithm}
\caption{The Proposed Fixed-Point Method for Problem $\mathcal{P}1.1.2$}
\begin{algorithmic}[1]
\State\textbf{Initialize:} Use superscript $(n)$ to denote the iteration number, give $t$ and $\mathbf{a}^{(0)}\in\mathcal{Q}$;
\State Set $n=1$;
\State \textbf{Repeat}
\State ~~Fix $\mathbf{a}^{(n-1)}$ and find the optimal $\{\mathbf{z}^{(n)}_{j}\}_{j\in\mathcal{K}\cup\mathcal{U}}$ using \eqref{SecIII-A-06} and \eqref{SecIII-A-07};
\State ~~Fix $\{\mathbf{z}^{(n)}_{j}\}_{j\in\mathcal{K}\cup\mathcal{U}}$ and apply the PGD method to the relaxed optimization problem
\begin{IEEEeqnarray}{ll}\nonumber
\mathcal{P}1.1.3:~\underset{\mathbf{a}^{(n)}}{\mathrm{max}}&~\sum^{G}_{i=1}\sum^{U}_{u=1}a_{i,u}\\
~~~~~~~~~~~~~~\mathrm{s.t.}&~\eqref{SecIII-A-02},\eqref{SecIII-A-01b},\eqref{SecIII-A-01d},\eqref{SecIII-A-05b},
\IEEEyesnumber
\IEEEyessubnumber\label{SecIII-A-12a}\\
&~0\leq a_{i,u}\leq 1,~\forall i\in\mathcal{G}, \forall u\in\mathcal{U},
\IEEEyesnumber
\IEEEyessubnumber\label{SecIII-A-12b}
\end{IEEEeqnarray}
~~according to \eqref{SecIII-A-10} and \eqref{SecIII-A-11};
\State \textbf{Until} convergence
\State Fix the optimized $\{\mathbf{z}^{*}_{j}\}_{j\in\mathcal{K}\cup\mathcal{U}}$ and apply BLP method to problem $\mathcal{P}1.1.3$ to find the optimal binary $\{a^{*}_{i,u}\}_{\forall i\in\mathcal{G}, \forall u\in\mathcal{U}}$;
\State \textbf{Return:} $\{a^{*}_{i,u}\}_{\forall i\in\mathcal{G}, \forall u\in\mathcal{U}}$ and $\{\mathbf{z}^{*}_{j}\}_{j\in\mathcal{K}\cup\mathcal{U}}$.
\end{algorithmic}
\end{algorithm}\\
The major computational complexity of \textit{Algorithm 1} lies in iteratively formulating beamforming vectors for each UE (i.e. Step 4) and the PGD method to find optimal association matrix (i.e. Step 5), and the BLP to finalize the optimal integer values of the association matrix (i.e. Step 7). Let us assume $L_{1}$ as the number of iterations to jointly optimize the beamforming vectors and association matrix. Thus, in each iteration, the worst-case complexity for optimizing the beamforming vectors is $\mathcal{O}((GK+U)N^{3})$, and the interior-point method for optimizing assocation matrix is $\mathcal{O}((GU)^{3}\log(1/\epsilon_{1}))$, where $\mathcal{O}(\log(1/\epsilon_{1}))$ is the complexity of interior-point iterations. In addition, following the work in \cite{Munapo2016}, our BLP problem can be converted into convex quadratic programming (CQP) problem, thus interior-point method can also be used to solve the problem in polynomial time. In summary, the worst-case computational complexity of \textit{Algorithm 1} is $\mathcal{O}(L_{1}(GK+U)N^{3}+(L_{1}+1)(GU)^{3}\log(1/\epsilon_{1}))$.

%%%%%%%%%%%%%%%%%%%%%%%%%%%%%%%%%%%%%%%
%%%%%%%%%%%%%%%%%%%%%%%%%%%%%%%%%%%%%%
\subsection{Inner Layer Iteration} 
Given the optimal UAV association matrix $\mathbf{A}$ from the outer layer iteration, in this inner layer iteration process, we want to jointly optimize UAVs' heights $h_{u}$, $\forall u\in\mathcal{U}$ and beamforming vectors $\mathbf{z}_{j}$, $\forall j\in\mathcal{K}\cup\mathcal{U}$. Specifically, let us denote $x_{u}\triangleq(h_u-h_{\mathrm{G}})^2$, then the UAVs' height related constraints can be formulated as
\begin{equation}\label{SecIII-B-01}
(h_{\mathrm{min}}-h_{\mathrm{G}})^2\leq x_{u}\leq \sum^{G}_{i=1}a_{i,u}\left[\left(\frac{\overline{P}_{\mathrm{U}}}{B_{\mathrm{L}}}\right)^{\frac{2}{\alpha_{l}}}-r^{2}_{i,u}\right],~\forall u\in\mathcal{U}.
\end{equation}
For ease of demonstration, we assume that the required transmit power of a UAV is always less than its maximum transmit power constraint $\overline{P}_{\mathrm{U}}$. Such assumption can be easily relaxed to more general cases if the association matrix $\mathbf{A}$ is pre-determined. Therefore, we have
\begin{IEEEeqnarray}{ll}\label{SecIII-B-01b}
P_{i,u}=\tilde{P}_{i,u}\triangleq a_{i,u}B_{\mathrm{L}}\cdot\left[r^{2}_{i,u}+x_{u}\right]^{\frac{\alpha_{l}}{2}},~\forall i\in\mathcal{G}, ~\forall u\in\mathcal{U}.
\end{IEEEeqnarray}
With the fixed $\mathbf{A}$, substituting $(h_u-h_{\mathrm{G}})^2$ with $x_{u}$ and substituting $P_{i,u}$ with $\tilde{P}_{i,u}$ for all equations in problem $\mathcal{P}1$, the sub-problem for inner layer iteration can be formulated as
\begin{IEEEeqnarray}{ll}\nonumber
\mathcal{P}1.2:\underset{\{\mathbf{z}_{j}\},\{x_{u}\}}{\mathrm{max}}&~t\nonumber\\
~~~~~~~~~~~~\mathrm{s.t.}&~\widetilde{\mathrm{SINR}}_{i,u}\geq a_{i,u}t,~\forall i\in\mathcal{G},~\forall u\in\mathcal{U},
\IEEEyesnumber
\IEEEyessubnumber\label{SecIII-B-02a}\\
&~\widetilde{\mathrm{SINR}}_{i,k}\geq \gamma,\forall i\in\mathcal{G},\forall k\in\mathcal{K}_{i},
\IEEEyesnumber
\IEEEyessubnumber\label{SecIII-B-02b}\\
&~\eqref{SecIII-B-01},
\IEEEyesnumber
\IEEEyessubnumber\label{SecIII-B-02c}
\end{IEEEeqnarray}
where we define
\begin{IEEEeqnarray}{ll}
\widetilde{\mathrm{SINR}}_{i,u}\triangleq\frac{a_{i,u}B_{\mathrm{L}}A_{\mathrm{L}}|\mathbf{z}^{H}_{u}\mathbf{h}_{i,u}|^2}{\sum_{k\in\mathcal{K}}\mathcal{D}^{(u)}_{i,k}+\sum_{g\in\mathcal{G}}\sum_{u^{'}\in\mathcal{U}/u}a_{g,u^{'}}\tilde{P}_{g,u^{'}}\mathcal{F}^{(u)}_{i,u^{'}}+\mathcal{N}^{(u)}_{i}},
\end{IEEEeqnarray}
and
\begin{IEEEeqnarray}{ll}
\widetilde{\mathrm{SINR}}_{i,k}\triangleq\frac{\mathcal{D}^{(k)}_{i,k}}{\sum_{k^{'}\in\mathcal{K}/k}\mathcal{D}^{(k)}_{i,k^{'}}+\sum_{g\in\mathcal{G}}\sum_{u\in\mathcal{U}}a_{g,u}\tilde{P}_{g,u}\mathcal{F}^{(k)}_{i,u}+\mathcal{N}^{(k)}_{i}}.
\end{IEEEeqnarray}
Problem $\mathcal{P}1.2$ can also be solved by using the generalized fixed-point method, which was used to solve problem $\mathcal{P}1.1.2$ in Sec. III-A. Specifically, given $x_{u}$, $\forall u\in\mathcal{U}$, each G-BS can compute the beamforming vectors for each of its associated UAV(s) as
\begin{IEEEeqnarray}{ll}\label{SecIII-B-03}
\mathbf{z}^{*}_{i,u}&=\arg\underset{\|\mathbf{z}_{i,u}\|=1}{\max}\frac{a_{i,u}B_{\mathrm{L}}A_{\mathrm{L}}\mathbf{z}^{H}_{i,u}\mathbf{h}_{i,u}\mathbf{h}^{H}_{i,u}\mathbf{z}_{i,u}}{\mathbf{z}^{H}_{i,u}\left(\sum_{k\in\mathcal{K}}\tilde{\mathcal{D}}_{i,k}+\sum_{g\in\mathcal{G}}\sum_{u^{'}\in\mathcal{U}/u}a_{g,u^{'}}\tilde{P}_{g,u^{'}}\tilde{\mathcal{F}}_{i,u^{'}}+\sigma^{2}_{0}\mathbf{I}\right)\mathbf{z}_{i,u}}\nonumber\\
&=a_{i,u}\eta_{i,u}\left(\sum_{k\in\mathcal{K}}\tilde{\mathcal{D}}_{i,k}+\sum_{g\in\mathcal{G}}\sum_{u^{'}\in\mathcal{U}/u}a_{g,u^{'}}\tilde{P}_{g,u^{'}}\tilde{\mathcal{F}}_{i,u^{'}}+\sigma^{2}_{0}\mathbf{I}\right)^{-1}\mathbf{h}_{i,u},~\forall i\in\mathcal{G},~\forall u\in\mathcal{U},
\end{IEEEeqnarray}
where $\eta_{i,u}$ denotes the normalization factor and is used to guarantee $\|\mathbf{z}_{i,u}\|=1$; $\tilde{\mathcal{D}}_{i,k}\triangleq P_{k}\zeta_{\mathrm{N}}(r_{i,k})\mathbf{h}_{i,k}\mathbf{h}^{H}_{i,k}$; $\tilde{\mathcal{F}}_{i,u^{'}}\triangleq\zeta_{\mathrm{L}}(r_{i,u^{'}})\mathbf{h}_{i,u^{'}}\mathbf{h}^{H}_{i,u^{'}}$. In addition, the beamforming vectors for each of G-BS associated G-UE can be formulated as
\begin{IEEEeqnarray}{ll}\label{SecIII-B-04}
\mathbf{z}^{*}_{i,k}&=\arg\underset{\|\mathbf{z}_{i,k}\|=1}{\max}\frac{P_{k}\zeta_{\mathrm{N}}(r_{i,k})\mathbf{z}^{H}_{i,k}\mathbf{h}_{i,k}\mathbf{h}^{H}_{i,k}\mathbf{z}_{i,k}}{\mathbf{z}^{H}_{i,k}\left(\sum_{k^{'}\in\mathcal{K}/k}\tilde{\mathcal{D}}_{i,k^{'}}+\sum_{g\in\mathcal{G}}\sum_{u\in\mathcal{U}}a_{g,u}\tilde{P}_{g,u}\tilde{\mathcal{F}}_{i,u}+\sigma^{2}_{0}\mathbf{I}\right)\mathbf{z}_{i,k}}\nonumber\\
&=\eta_{i,k}\left(\sum_{k^{'}\in\mathcal{K}/k}\tilde{\mathcal{D}}_{i,k^{'}}+\sum_{g\in\mathcal{G}}\sum_{u\in\mathcal{U}}a_{g,u}\tilde{P}_{g,u}\tilde{\mathcal{F}}_{i,u}+\sigma^{2}_{0}\mathbf{I}\right)^{-1}\mathbf{h}_{i,k},~\forall i\in\mathcal{G},~\forall k\in\mathcal{K}_{i},
\end{IEEEeqnarray}
where $\eta_{i,k}$ denotes the normalization factor and is used to guarantee $\|\mathbf{z}_{i,k}\|=1$.

Given the above formulated beamforming vectors, the optimal UAVs' height can be calculated as follow. As we can see, the second term in the denominator of SINR in \eqref{SecIII-B-02a} can be expanded to
\begin{IEEEeqnarray}{ll}\label{Add_01}
\sum_{g\in\mathcal{G}}\sum_{u^{'}\in\mathcal{U}/u}a_{g,u^{'}}\tilde{P}_{g,u^{'}}\mathcal{F}^{(u)}_{i,u^{'}}=&
\sum_{u^{'}\in\mathcal{U}_{i}/u}a_{i,u^{'}}B_{\mathrm{L}}A_{\mathrm{L}}|\mathbf{z}^{H}_{i,u}\mathbf{h}_{i,u^{'}}|^{2}\nonumber\\
&+\sum_{g\in\mathcal{G}/i}\sum_{u^{'}\in\mathcal{U}/\mathcal{U}_{i}}a_{g,u^{'}}B_{\mathrm{L}}A_{\mathrm{L}}|\mathbf{z}^{H}_{u}\mathbf{h}_{i,u^{'}}|^{2}\left(\frac{r^{2}_{g,u^{'}}+x_{u^{'}}}{r^{2}_{i,u^{'}}+x_{u^{'}}}\right)^{\frac{\alpha_{l}}{2}},
\end{IEEEeqnarray}
where the first term in the right-hand side of \eqref{Add_01} is the intra-cell interference, and the second term in the right-hand side of \eqref{Add_01} is the inter-cell interference. Similarly, the second term in the denominator of SINR in \eqref{SecIII-B-02b} can be expanded to
\begin{IEEEeqnarray}{ll}\label{Add_02}
\sum_{g\in\mathcal{G}}\sum_{u\in\mathcal{U}}a_{g,u}\tilde{P}_{g,u}\mathcal{F}^{(k)}_{i,u}=&
\sum_{u\in\mathcal{U}_{i}}a_{i,u}B_{\mathrm{L}}A_{\mathrm{L}}|\mathbf{z}^{H}_{k}\mathbf{h}_{i,u}|^{2}+\sum_{g\in\mathcal{G}/i}\sum_{u\in\mathcal{U}/\mathcal{U}_{i}}a_{g,u}B_{\mathrm{L}}A_{\mathrm{L}}|\mathbf{z}^{H}_{k}\mathbf{h}_{i,u}|^{2}\left(\frac{r^{2}_{g,u}+x_{u}}{r^{2}_{i,u}+x_{u}}\right)^{\frac{\alpha_{l}}{2}}, \nonumber\\
\end{IEEEeqnarray}
where the first term in the right-hand side of \eqref{Add_02} is the intra-cell interference, and the second term in the right-hand side of \eqref{Add_02} is the inter-cell interference.

Let us define $\hat{\mathcal{F}}^{(j)}_{i,u}\triangleq B_{\mathrm{L}}A_{\mathrm{L}}|\mathbf{z}^{H}_{j}\mathbf{h}_{i,u}|^2$, $\mathcal{I}^{(u)}_{i}\triangleq\sum_{k\in\mathcal{K}}\mathcal{D}^{(u)}_{i,k}+\sum_{u^{'}\in\mathcal{U}_{i}/u}a_{i,u^{'}}B_{\mathrm{L}}A_{\mathrm{L}}|\mathbf{z}^{H}_{u}\mathbf{h}_{i,u^{'}}|^{2}$ as the sum of intra-cell interference at the $i^{\mathrm{th}}$ G-BS for the $u^{\mathrm{th}}$ UAV, and $\mathcal{I}^{(k)}_{i}\triangleq\sum_{k^{'}\in\mathcal{K}/k}\mathcal{D}^{(k)}_{i,k^{'}}+\sum_{u\in\mathcal{U}_{i}}a_{i,u}B_{\mathrm{L}}A_{\mathrm{L}}|\mathbf{z}^{H}_{k}\mathbf{h}_{i,u}|^{2}$ as the sum of intra-cell interference at the $i^{\mathrm{th}}$ G-BS for the $k^{\mathrm{th}}$ G-UE. Through introducing the auxiliary variables $x^{u}_{i,g}$, $\forall i, g\neq i, u$, the problem $\mathcal{P}1.2$ with fixed $\mathbf{z}_{i,u}$ and $\mathbf{z}_{i,k}$, $\forall i,u,k,$ can be reformulated as
\begin{IEEEeqnarray}{ll}\nonumber
\mathcal{P}1.2.1:~\underset{\{x_{u}\},\{x^{u}_{i,g}\}}{\mathrm{max}}&~t\\
~~~~~~~~~~~~~~~~\mathrm{s.t.}&~\widehat{\mathrm{SINR}}_{i,u}\geq a_{i,u}t,\forall i\in\mathcal{G}, \forall u\in\mathcal{U},
\IEEEyesnumber
\IEEEyessubnumber\label{SecIII-B-05a}\\
&~\widehat{\mathrm{SINR}}_{i,k}\geq \gamma,\forall i\in\mathcal{G},\forall k\in\mathcal{K}_{i},
\IEEEyesnumber
\IEEEyessubnumber\label{SecIII-B-05b}\\
&~\left(\frac{r^{2}_{g,u}+x_{u}}{r^{2}_{i,u}+x_{u}}\right)^{\frac{\alpha_{l}}{2}}\leq x^{u}_{i,g},~\forall i,g\neq i\in\mathcal{G},~\forall u\in\mathcal{U},
\IEEEyesnumber
\IEEEyessubnumber\label{SecIII-B-05c}\\
&~\eqref{SecIII-B-01},
\IEEEyesnumber
\IEEEyessubnumber\label{SecIII-B-05d}
\end{IEEEeqnarray}
where we define
\begin{IEEEeqnarray}{ll}
\widehat{\mathrm{SINR}}_{i,u}\triangleq\frac{a_{i,u}\hat{\mathcal{F}}^{u}_{i,u}}{\mathcal{I}^{(u)}_{i}+\sum_{g\in\mathcal{G}/i}\sum_{u^{'}\in\mathcal{U}/\mathcal{U}_{i}}a_{g,u^{'}}\hat{\mathcal{F}}^{(u)}_{i,u^{'}}x^{u^{'}}_{i,g}+\mathcal{N}^{(u)}_{i}},
\end{IEEEeqnarray}
and
\begin{IEEEeqnarray}{ll}
\widehat{\mathrm{SINR}}_{i,k}\triangleq\frac{\mathcal{D}^{(k)}_{i,k}}{\mathcal{I}^{(k)}_{i}+  \sum_{g\in\mathcal{G}/i}\sum_{u\in\mathcal{U}/\mathcal{U}_{i}}a_{g,u}\hat{\mathcal{F}}^{(k)}_{i,u}x^{u}_{i,g}+\mathcal{N}^{(k)}_{i}}.
\end{IEEEeqnarray}
Problem $\mathcal{P}1.2.1$ is still non-convex due to the coupling between $x^{u^{'}}_{i,g}$ and $t$ in \eqref{SecIII-B-05a} and the non-convexity on the left-hand side in \eqref{SecIII-B-05c}. Inspired by the GGP modelling as in \cite{Boyd2007}, the following lemma will help us to convert the non-convex problem $\mathcal{P}1.2.1$ into a convex problem.

{\em Lemma 1:} If a function $f(x)$ is a posynomial, i.e., 
\begin{equation}\label{SecIII-B-06}
f(x)=\sum^{K}_{k=1}c_{k}x^{a_{1k}}_{1}x^{a_{2k}}_{2}\cdots x^{a_{nk}}_{n},
\end{equation}
where $c_{k}>0$ is constant, $a_{i}\in\mathbb{R}$ is exponents, and $x_{1},x_{2},\cdots,x_{n}$ are variables. By defining $y_{i}=\log x_{i}$ (so $x_{i}=e^{y_{i}}$), the function 
\begin{equation}\label{SecIII-B-07}
F(y)=\log f(e^{y})
\end{equation}
is convex \cite{Boyd2007}.

According to {\em Lemma 1}, constraint \eqref{SecIII-B-05a} is a standard posynomial inequality. By defining $y^{u}_{i,g}\triangleq\log x^{u}_{i,g}$ (so $x^{u}_{i,g}=e^{y^{u}_{i,g}}$), $\forall i,g\neq i\in\mathcal{G}, \forall u\in\mathcal{U}$, and $y_{t}\triangleq \log t$ (so $t=e^{y_{t}}$), constraint \eqref{SecIII-B-05a} can be converted to a convex constraint, which is
\begin{IEEEeqnarray}{ll}\label{SecIII-B-08}
\frac{a_{i,u}\mathcal{I}^{(u)}_{i}}{a_{i,u}\hat{\mathcal{F}}^{u}_{i,u}}\cdot e^{y_{t}}+\sum_{g\in\mathcal{G}/i}\sum_{u^{'}\in\mathcal{U}/\mathcal{U}_{i}}\frac{a_{i,u}a_{g,u^{'}}\hat{\mathcal{F}}^{(u)}_{i,u^{'}}}{a_{i,u}\hat{\mathcal{F}}^{u}_{i,u}}\cdot e^{y^{u^{'}}_{i,g}}\cdot e^{y_{t}}+\frac{a_{i,u}\mathcal{N}^{(u)}_{i}}{a_{i,u}\hat{\mathcal{F}}^{u}_{i,u}}\cdot e^{y_{t}}-1\leq 0.
\end{IEEEeqnarray}
Similarly, constraint \eqref{SecIII-B-05b} is also a standard posynomial inequality, and in order to in line with \eqref{SecIII-B-08}, we have
\begin{IEEEeqnarray}{ll}\label{SecIII-B-09}
\frac{\gamma\mathcal{I}^{(k)}_{i}}{\mathcal{D}^{(k)}_{i,k}}+\sum_{g\in\mathcal{G}/i}\sum_{u\in\mathcal{U}/\mathcal{U}_{i}}\frac{\gamma a_{g,u}\hat{\mathcal{F}}^{(k)}_{i,u}}{\mathcal{D}^{(k)}_{i,k}}e^{y^{u}_{i,g}}+\frac{\gamma\mathcal{N}^{(k)}_{i}}{\mathcal{D}^{(k)}_{i,k}}-1\leq 0.
\end{IEEEeqnarray}
In addition, by introducing logarithm operation and then replacing the concave part of left-hand side of constraint \eqref{SecIII-B-05c} by its first order Taylor expansion, the constraint \eqref{SecIII-B-05c} can be converted into
\begin{equation}\label{SecIII-B-10}
\log(r^{2}_{g,u}+\overline{x}_{u})+\frac{1}{r^{2}_{g,u}+\overline{x}_{u}}(x_u-\overline{x}_u)-\log(r^{2}_{i,u}+x_{u})\leq \frac{2}{\alpha_{l}}y^{u}_{i,g},
\end{equation}
where $\overline{x}_{u}$ is the optimal solution obtained from the previous iteration. Based on the CCP as in \cite{Tao2016}, the solution of non-convex problem $\mathcal{P}1.2.1$ can be obtained by successively solving a sequence of the following convex problem:
\begin{IEEEeqnarray}{ll}\nonumber
\mathcal{P}1.2.2:~\underset{y_t,\{x_{u}\}\{y^{(u)}_{i,g}\}}{\mathrm{max}}&~y_{t}\\
~~~~~~~~~~~~~~~~~~~~~~\mathrm{s.t.}&~\eqref{SecIII-B-01},\eqref{SecIII-B-08},\eqref{SecIII-B-09},\eqref{SecIII-B-10},
\IEEEyesnumber
\IEEEyessubnumber\label{SecIII-B-11a}
\end{IEEEeqnarray}
where the standard interior point method \cite{Boyd2004} can be used to solve the problem $\mathcal{P}1.2.2$.

The overall steps to solve problem $\mathcal{P}1.2$ is given in \textit{Algorithm 2} in detail. 
\begin{algorithm}
\caption{The Proposed Fixed-Point Method for Problem $\mathcal{P}1.2$}
\begin{algorithmic}[1]
\State\textbf{Initialize:} Use superscript $(n)$ to denote the iteration number, give $\mathbf{A}$, and set $h^{(0)}_{u}=h_{\mathrm{min}},\forall u\in\mathcal{U}$;
\State Set $n=1$;
\State \textbf{Repeat}
\State ~~Fix $\{h^{(n-1)}_{u}\}_{\forall u\in\mathcal{U}}$ and find the optimal $\{\mathbf{z}^{(n)}_{i,u}\}_{\forall i\in\mathcal{G},\forall u\in\mathcal{U}}$ and $\{\mathbf{z}^{(n)}_{i,k}\}_{\forall i\in\mathcal{G},\forall k\in G-UE_{i}}$ using \eqref{SecIII-B-03}\\ ~~and \eqref{SecIII-B-04}, respectively;
\State ~~Fix $\{\mathbf{z}^{(n)}_{i,u}\}_{\forall i\in\mathcal{G},\forall u\in\mathcal{U}}$ and $\{\mathbf{z}^{(n)}_{i,k}\}_{\forall i\in\mathcal{G},\forall k\in G-UE_{i}}$ and apply CCP at starting point $\overline{x}^{(n-1)}_{u}=$\\ ~~$(h^{(n-1)}_{u}-h_{\mathrm{G}})^2, \forall u\in\mathcal{U}$ to solve problem $\mathcal{P}1.2.2$ and obtain the solution $\{h^{(n)}_{u}\}_{\forall u\in\mathcal{U}}$;
\State \textbf{Until} convergence
\State \textbf{Return:} $\{h^{*}_{u}\}_{\forall u\in\mathcal{U}}$, $\{\mathbf{z}^{*}_{i,u}\}_{\forall i\in\mathcal{G},\forall u\in\mathcal{U}}$ and $\{\mathbf{z}^{*}_{i,k}\}_{\forall i\in\mathcal{G},\forall k\in G-UE_{i}}$.
\end{algorithmic}
\end{algorithm}\\
The major computational complexity of \textit{Algorithm 2} lies in the minimum mean square error based beamforming vector formulation and successive CCP for optimizing UAVs' height per outer iteration. Let us assume the number of outer fixed-point iterations as $L_{2}$ and the number of iterations for successive CCP as $L_{3}$. In each outer iteration, the main computational complexity for optimizing beamforming vectors is $\mathcal{O}((GK+U)^{3}$, and the complexity for optimizing UAVs' height with CCP is $\mathcal{O}(L_{3}(U+V+1)^{3}\log(1/\epsilon_{1}))$, where $V$ denotes the number of random variables of $x^{u}_{i,g}, \forall i,g\neq i\in\mathcal{G},\forall u\in\mathcal{U}$. Thus, the worst-case computational complexity of \textit{Algorithm 2} is $\mathcal{O}(L_{2}(GK+U)^{3}+L_{2}L_{3}(U+V+1)^{3}\log(1/\epsilon_{1}))$.

%%%%%%%%%%%%%%%%%%%%%%%%%%%%%%%%%%%%%%5
\subsection{Algorithm Outline and Convergence Analysis}
\subsubsection{Algorithm Outline}
Combining the joint UAV association matrix $\mathbf{A}$ and beamforming vectors optimization in Sec. III-A and the joint UAVs' height and beamforming vectors optimization in Sec. III-B, the proposed hierarchical bi-layer search method to solve problem $\mathcal{P}1$ can be outlined in following \textit{Algorithm 3}. 
\begin{algorithm}
\caption{The Proposed Bi-Layer Search Method for Problem $\mathcal{P}1$} 
\begin{algorithmic}[1]
\State\textbf{Initialize:} Use superscript $(n)$ and $(m)$ to denote the iteration numbers, give $t^{(0)}=\frac{\underline{t}+\overline{t}}{2}$, and set $h^{(0)}_{u}=h_{\mathrm{min}},\forall u\in\mathcal{U}$;
\State Set $n=1$;
\State \textbf{Repeat}
\State ~~Give $\{h^{(n-1)}\}_{\forall u\in\mathcal{U}}$ and set $m=1$;
\State ~~\textbf{Repeat}
\State ~~~~Give $t^{(m-1)}$, solve problem $\mathcal{P}1.1.2$ using \textit{Algorithm 1}, and identify the problem $\mathcal{P}1.1.2$\\~~~~`feasible' or not;
\State ~~~~Update $t^{(m)}$ by using \eqref{SecIII-A-03} or \eqref{SecIII-A-04};
\State ~~\textbf{Until} convergence
\State ~~Give $\mathbf{A}^{(n)}$, solve problem $\mathcal{P}1.2$ using \textit{Algorithm 2} and obtain $\{h^{(n)}_{u}\}_{\forall u\in\mathcal{U}}$;
\State \textbf{Until} convergence
\State \textbf{Return:} $\mathbf{A}^{*}$, $\{h^{*}_{u}\}_{\forall u\in\mathcal{U}}$, $\{\mathbf{z}^{*}_{i,u}\}_{\forall i\in\mathcal{G},\forall u\in\mathcal{U}}$ and $\{\mathbf{z}^{*}_{i,k}\}_{\forall i\in\mathcal{G},\forall k\in G-UE_{i}}$.
\end{algorithmic}
\end{algorithm}\\
By involving the computational complexity of \textit{Algorithm 1} and \textit{Algorithm 2}, the polynomial time computational complexity for solving problem $\mathcal{P}1$ is $\mathcal{O}(L_{4}(L_{1}(GK+U)N^{3}+(L_{1}+1)(GU)^{3}\log(1/\epsilon_{1}))\log(1/\epsilon_{2})+L_{4}(L_{2}(GK+U)^{3}+L_{2}L_{3}(U+V+1)^{3}\log(1/\epsilon_{1})))$, where $L_{4}$ denotes the number of iterations of the proposed bi-layer search method, and $\mathcal{O}(\log(1/\epsilon_{2}))$ is the complexity of the bisection method involved in the outer layer iteration.% It is worth noting that problem $\mathcal{P}2$ can be solved using a modified version of \textit{Algorithm 3}, where the modification is to remove the bi-section search process in \textit{Algorithm 3}.  

\subsubsection{Convergence of the Inner Layer Iteration}
To show the convergence of the inner layer iteration, we first introduce the following Proposition to show that $f(t)$, representing the maximum number of active UAVs, is strictly monotonically decreasing.

{\em Proposition 1:} For any given feasible target SINRs for both UAVs and G-UEs, the maximum number of activated UAVs $f(t)$ is strictly monotonically decreasing with respect to the target SINR, i.e., $t$, of UAVs. 
\begin{proof}
The monotonicity can be verified by contradiction. Assume that by given some values $t>\tilde{t}$, we have $f(t)\geq f(\tilde{t})$. Their solutions for problem $\mathcal{P}1.1.2$ are denoted as $\{\mathbf{a}^{*},\{\mathbf{z}^{*}_{i,j}\}_{i\in\mathcal{G},j\in\mathcal{K}\cup\mathcal{U}}\}$ and $\{\tilde{\mathbf{a}}^{*},\{\tilde{\mathbf{z}}^{*}_{i,j}\}_{i\in\mathcal{G},j\in\mathcal{K}\cup\mathcal{U}}\}$, respectively, where $\mathbf{a}=[a_{1,1},\ldots,a_{G,U}]^{T}$ is the column-wise association matrix $\mathbf{A}$. From \eqref{SecIII-A-02} and \eqref{SecIII-A-01b}, we can observe that $\mathrm{SINR}(c\mathbf{a},\mathbf{z}_{i,j})<\mathrm{SINR}(\mathbf{a},\mathbf{z}_{i,j})$ for all $c>1$, which means we can always find a $c_{0}>1$ such that $\mathrm{SINR}(c_{0}\mathbf{a},\mathbf{z}_{i,j})=\tilde{t}$. In this case, considering the assumption that $f(t)\geq f(\tilde{t})$, we infer $\|c_{0}\mathbf{a}\|_{1}>\|\mathbf{a}\|_{1}\geq\|\tilde{\mathbf{a}}\|_{1}$, which means that $\{\tilde{\mathbf{a}}^{*},\{\tilde{\mathbf{z}}^{*}_{i,j}\}_{i\in\mathcal{G},j\in\mathcal{K}\cup\mathcal{U}}\}$ is not optimal for $\tilde{t}$ and contradicts the original assumption.
\end{proof}

With \textit{Proposition 1}, we can first conclude that the bi-section method used in outer iteration in Sec. III-A-1 can always guarantee the convergence. Then, for the inner iteration in Sec. III-A-2, it is clear that in the steps of updating the UAV association matrix $\mathbf{A}$, the objective function of problem $\mathcal{P}1.1.2$ is maximized while the beamforming vector $\{\mathbf{z}_{i,j}\}_{i\in\mathcal{G},j\in\mathcal{K}\cup\mathcal{U}}$ are held fixed. On the other hand, with fixed $\mathbf{A}$ and assisted by \textit{Proposition 1}, finding the optimal beamforming vectors to maximize the actual SINR will weaken the restriction of the target SINR, which leave more DoFs to be used to further increase the number of active UAVs. Then, by following the well-known fixed-point iteration theory \cite{Yates1995}, the convergence of the inner iteration (i.e., \textit{Algorithm 1}) can be guaranteed as well.    

\subsubsection{Convergence of the Outer Layer Iteration}
By inserting the optimal beamforming vectors from \eqref{SecIII-B-03} into the left-hand side of SINR constraints of UAVs in \eqref{SecIII-B-05a}, we obtain
\begin{IEEEeqnarray}{ll}
\mathrm{SINR}_{i,u} = a_{i,u}B_{\mathrm{L}}A_{\mathrm{L}}\mathbf{h}^{H}_{i,u}(\mathcal{I}^{(u)}_{i}+\sum_{g\in\mathcal{G}/i}\sum_{u^{'}\in\mathcal{U}/\mathcal{U}_{i}}a_{g,u^{'}}B_{\mathrm{L}}A_{\mathrm{L}}\mathbf{h}^{H}_{i,u^{'}}\mathbf{h}_{i,u^{'}}x^{u^{'}}_{i,g}+\mathcal{N}^{(u)}_{i})^{-1}\mathbf{h}_{i,u},
\end{IEEEeqnarray}
which is a continuously and monotonously decreasing function with respect to $\{x^{u^{'}}_{i,g}\}$. In this case, by iteratively updating $\{x^{u^{'}}_{i,g}\}$ within its feasibility region, we can always find a unique fixed point, i.e., $\{x^{u^{'*}}_{i,g}\}$, that lead to the maximum target SINR $t^{*}$ of UAVs. Thus, convergence of outer iterations of \textit{Algorithm 2} can be guaranteed. On the other hand, the convergence proof of CCP method to iteratively optimize $\{x^{u^{'}}_{i,g}\}$ and $\{x_{u}\}$ in \textit{Algorithm 2} can be found in the work \cite{Tao2016}. In summary, with the guaranteed convergence of inner layer and outer layer iterations, the convergence of the proposed \textit{Algorithm 3} for solving the original problem $\mathcal{P}1$ can be easily guaranteed with the help of the fixed-point iteration theory \cite{Yates1995}.

%%%%%%%%%%%%%%%%%%%%%%%%%%%%%%%%%%%%%%%%%%%
\section{Robust Proposed Algorithm with Imperfect CSI}
One of the key challenges in cellular-connected UAV communication systems is how to ensure efficient mobility performances. In this case, perfectly tracking the rapidly changed channels is almost impossible. Moreover, due to emerging trends in small-cell, the size of G-BS is getting smaller. With the fixed number of G-BS antennas, if they are not well separated by more than the coherence distance, different channel links will experience spatial correlations as well. Motivated by these observations, in this section, we examine the proposed algorithm in Sec. III considering statistical channel environments. In this case, the assumed imperfect CSI consists of the first and second order statistics of the actual channel, i.e., the channel is modelled as the estimated version plus its estimation error covariance. To this end, such channel model can be mathematically formulated as \cite{Zhang2008,Xiao2010}
\begin{equation}\label{SecIV-01}
\mathbf{h}_{i,j}=\overline{\mathbf{h}}_{i,j}+\mathbf{R}^{1/2}_{i}\mathbf{h}_{w},~\forall i\in\mathcal{G},~\forall j\in\mathcal{K}\cup\mathcal{U},
\end{equation}
where $\mathbf{h}_{w}\in\mathbb{C}^{N\times1}$ denotes the channel estimation error, and has independent and identically distributed (i.i.d.) elements distributed as $\mathcal{CN}(0,\sigma^{2}_{\mathrm{ch}})$. In \eqref{SecIV-01}, $\mathbf{R}_{i}$ is defined as the long-term receive correlation matrix at the $i^{\mathrm{th}}$ G-BS, which can be formulated as a Toeplitz matrix \cite{Kiessling2004}, defined by the correlation coefficient $\rho$ ($0\leq\rho\leq1$) as $[\mathbf{R}_{i}]_{m,n}=\rho^{|m-n|}, \forall i\in\mathcal{G}$.

While the actual performance metric of communication systems, such as data rate or bit-error-rate, is normally as a function of SINR. 
When G-BSs have perfect CSI, the aforementioned metrics are all function of the same SINR as it defined. However, if imperfect CSI is assumed, it is harder to measure system performances with direct link to the SINR. This motivates us to focus on the ergodic capacity lower bound of the system to rigorously characterize the system performances. It is worth noting that the exact capacity is unknown in general. In light of the work in \cite{Sanguinetti2019}, we use one popular choice of capacity lower bound, named \textit{user-and-then-forget (UatF) bound}\cite{Bjornson2017}, to measure the system performance, which is
\begin{IEEEeqnarray}{ll}\label{SecIV-02}
{R}_{i,j}\triangleq\log_2\left(1+\overline{\mathrm{SINR}}_{i,j}\right),~\forall i\in\mathcal{G},~\forall j\in\mathcal{K}\cup\mathcal{U},
\end{IEEEeqnarray}
and $\overline{\mathrm{SINR}}_{i,j}$ is denoted as the \textit{effective} SINR. For G-UEs, we have
\begin{IEEEeqnarray}{ll}\label{SecIV-03}
\overline{\mathrm{SINR}}_{i,k}&=\frac{P_{k}\zeta_{\mathrm{N}}(r_{i,k})|\mathbb{E}\{\mathbf{z}^{H}_{k}\mathbf{h}_{i,k}\}|^2}{\sum_{k\in\mathcal{K}}\mathbb{E}\{\mathcal{D}^{(k)}_{i,k}\}+\sum_{g\in\mathcal{G}}\sum_{u\in\mathcal{U}}a_{g,u}P_{g,u}\mathbb{E}\{\mathcal{F}^{(k)}_{i,u}\}-P_{k}\zeta_{\mathrm{N}}(r_{i,k})|\mathbb{E}\{\mathbf{z}^{H}_{k}\mathbf{h}_{i,k}\}|^2+\sigma^{2}_{0}},\nonumber\\ 
\end{IEEEeqnarray}
and for UAVs, we have
\begin{IEEEeqnarray}{ll}\label{SecIV-04}
\overline{\mathrm{SINR}}_{i,u} = \frac{a_{i,u}P_{i,u}\zeta_{\mathrm{L}}(r_{i,u})|\mathbb{E}\{\mathbf{z}^{H}_{u}\mathbf{h}_{i,u}\}|^2}{\sum_{k\in\mathcal{K}}\mathbb{E}\{\mathcal{D}^{(u)}_{i,k}\}+\sum_{g\in\mathcal{G}}\sum_{u\in\mathcal{U}}a_{g,u}P_{g,u}\mathbb{E}\{\mathcal{F}^{(u)}_{i,u}\}-a_{i,u}P_{i,u}\zeta_{\mathrm{L}}(r_{i,u})|\mathbb{E}\{\mathbf{z}^{H}_{u}\mathbf{h}_{i,u}\}|^2+\sigma^{2}_{0}},\nonumber\\
\end{IEEEeqnarray}
where the expectations $\mathbb{E}\{\cdot\}$ in \eqref{SecIV-03} and \eqref{SecIV-04} are taken with respect to the small-scale fading channel $\mathbf{h}_{i,j}$, $\forall i\in\mathcal{G}, \forall j\in\mathcal{K}\cup\mathcal{U}$. Following the imperfect CSI model in \eqref{SecIV-02}, we have
\begin{IEEEeqnarray}{ll}\label{SecIV-05}
|\mathbb{E}\{\mathbf{z}^{H}_{j}\mathbf{h}_{i,j}\}|^2&=\mathbf{z}^{H}_{j}\mathbb{E}\{(\overline{\mathbf{h}}_{i,j}+\mathbf{R}^{1/2}_{i}\mathbf{h}_{w})\}\mathbb{E}\{(\overline{\mathbf{h}}^{H}_{i,j}+\mathbf{h}^{H}_{w}\mathbf{R}^{H/2}_{i})\}\mathbf{z}^{H}_{j}\nonumber\\
&=\mathbf{z}^{H}_{j}(\overline{\mathbf{h}}_{i,j}+\mathbb{E}\{\mathbf{R}^{1/2}_{i}\mathbf{h}_{w}\})(\overline{\mathbf{h}}^{H}_{i,j}+\mathbb{E}\{\mathbf{h}^{H}_{w}\mathbf{R}^{H/2}_{i}\})\mathbf{z}^{H}_{j}\nonumber\\
&=\mathbf{z}^{H}_{j}\overline{\mathbf{h}}_{i,j}\overline{\mathbf{h}}^{H}_{i,j}\mathbf{z}_{j},
\end{IEEEeqnarray}
and
\begin{IEEEeqnarray}{ll}\label{SecIV-06}
\mathbb{E}\{|\mathbf{z}^{H}_{j}\mathbf{h}_{i,j}|^2\}&=\mathbf{z}^{H}_{j}\mathbb{E}\{(\overline{\mathbf{h}}_{i,j}+\mathbf{R}^{1/2}_{i}\mathbf{h}_{w})(\overline{\mathbf{h}}^{H}_{i,j}+\mathbf{h}^{H}_{w}\mathbf{R}^{H/2}_{i})\}\mathbf{z}^{H}_{j}\nonumber\\
&=\mathbf{z}^{H}_{j}(\overline{\mathbf{h}}_{i,j}\overline{\mathbf{h}}^{H}_{i,j}+\mathbb{E}\{\mathbf{R}^{1/2}_{i}\mathbf{h}_{w}\mathbf{h}^{H}_{w}\mathbf{R}^{H/2}_{i}\})\mathbf{z}^{H}_{j}\nonumber\\
&=\mathbf{z}^{H}_{j}\overline{\mathbf{h}}_{i,j}\overline{\mathbf{h}}^{H}_{i,j}\mathbf{z}_{j}+\sigma^{2}_{\mathrm{ch}}\mathbf{z}^{H}_{j}\mathbf{R}_{i}\mathbf{z}_{j}.
\end{IEEEeqnarray}
Then, by inserting \eqref{SecIV-05} and \eqref{SecIV-06} into \eqref{SecIV-03} and \eqref{SecIV-04}, and replacing the SINR constraints in \eqref{SecII-C-01a} and \eqref{SecII-C-01b} with \eqref{SecIV-03} and \eqref{SecIV-04}, respectively, the proposed algorithm that was used to solve problem $\mathcal{P}1$ in Sec. III can be directly used to solve the newly formulated objective problem with consideration of imperfect CSI. It is worth noting that, due to the mobility of UAVs, we have to update our optimized UAV association matrix, beamforming vectors, and/or height of the UAVs when a period of time lapses. More frequent updates will lead to more accurate system evaluations. However, more updates will also bring the increased computational complexity. In the next section, computer simulation will be used to evaluate the trade-off between UAVs' rate performances and the frequency to update the optimization variables.

\section{Numerical Results and Discussion}
In this section, comprehensive simulations are provided to illustrate the performance of the proposed joint beamforming, UAV association, and power control algorithm under both perfect and imperfect CSI. Consider that three G-BSs (i.e., $G=3$) are evenly distributed in a square area of the size $600\text{m} \times 600\text{m}$ to serve both UAVs and their pre-associated randomly distributed G-UEs. The height of each G-BS is $h_{\mathrm{G}}=25\text{m}$, and the noise power spectrum density is $\sigma^{2}_{0}=-174\mathrm{dBm/Hz}$. The channel path loss between UAV and G-BS is $A_{\mathrm{L}}=6\times 10^{-3}$, and the path loss exponent is $\alpha_{\mathrm{L}}=2.09$. Correspondingly, the channel path loss between G-UE and G-BS is $A_{\mathrm{N}}=10^{-3}$, and the path loss exponent is $\alpha_{\mathrm{N}}=3.75$. The small-scale fading for the UAV to G-BS channel links follows Nakagami-3 distribution and the small-scale fading for G-UE to G-BS channel links follows Nakagami-1 distribution. Following 3GPP standard \cite{3GPP2017}, both UAVs and G-UEs implement uplink power control with cell-specific parameters $B_{\mathrm{L}}=1.2\times 10^{-8}$ and $B_{\mathrm{L}}=1.67\times 10^{-9}$, respectively. Throughout the simulations, 1000 channel realizations are conducted to compute the average UAV's/G-UE's rate performances. 
%%%%%%%%%%%%%%%%%%%%%%%%%%%
\subsection{Achievable Rate Versus Number of G-BS Antennas}
\begin{figure}[t] \setlength{\belowcaptionskip}{-.5cm}
\begin{center}
\epsfig{figure=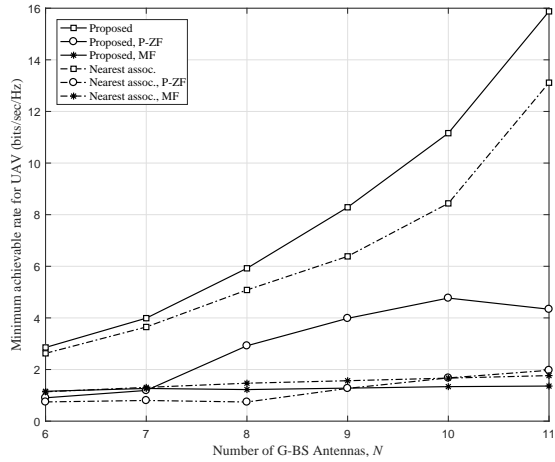,scale=0.44,angle=0}
\end{center}
\caption{Minimum achievable rate for UAV versus number of G-BS antennas.}\label{F1}
\end{figure}

\begin{figure}[t] \setlength{\belowcaptionskip}{-.5cm}
\begin{center}
\epsfig{figure=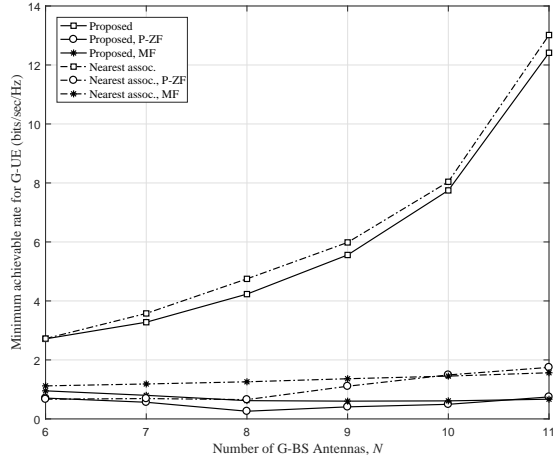,scale=0.44,angle=0}
\end{center}
\caption{Minimum achievable rate for G-UE versus number of G-BS antennas.}\label{F2}
\end{figure}

Considering perfect CSI estimation, in this subsection we examine the average minimum achievable rate of our proposed optimization algorithm in comparison to the conventional UAV association via the nearest G-BS (e.g., nearest association) for various number of G-BS antennas. We set the number of G-UEs per cell as $K=2$ and the total number of UAVs as $U=6$. The target G-UE's SINR is set to be $\gamma=2$. The UAVs' minimum achievable rate is illustrated in Fig.~\ref{F1}.
In comparison to the nearest UAV association method, our proposed algorithm achieves higher UAV's rate especially when the number of G-BS antennas increases. This is because, with increasing the number of G-BS antenna, the UAVs have more degrees of freedom (DoFs) to choose their preferred G-BSs, which results in higher UAV's rate in our proposed algorithm. Furthermore, we also plot Fig.~\ref{F2} to show the minimum achievable rate for G-UE.
It can be seen that, although the achievable rate for our proposed algorithm is slightly smaller than that of the nearest association case, the minimum target rate for G-UE can still be guaranteed for both cases, e.g., $\log_{2}(1+\gamma)\geq 1.58~\text{bits/sec/Hz}$. 

In addition, given the optimized UAV association matrix and the height of UAVs obtained from \textit{Algorithm 3}, we also plot the minimum achievable rates with partial zero-forcing (P-ZF) \cite{Jindal2011} and match-filtering (MF) based beamforming designs in both Fig.~\ref{F1} and Fig.~\ref{F2}. For P-ZF beamforming, each G-BS needs to cancel all intra-cell interference and inter-cell interference from UAV(s), and if there is still DoF(s) left, the G-BS will use the remaining DoF(s) to boost the power of its desired signal. It is shown that the P-ZF with the proposed UAV association method can still outperforms the nearest UAV association case in terms of UAVs' minimum achievable rate. For other cases, the nearest UAV association method will lead to better minimum rate performance. Importantly, our proposed optimal beamforming strategy always leads to superior minimum rate performances than P-ZF and MF beamforming methods.
%%%%%%%%%%%%%%%%%%%%%%%%
\subsection{Achievable Rate Versus Height of UAVs, and Number of G-UEs}
\begin{figure}[t] \setlength{\belowcaptionskip}{-.5cm}
\begin{center}
\epsfig{figure=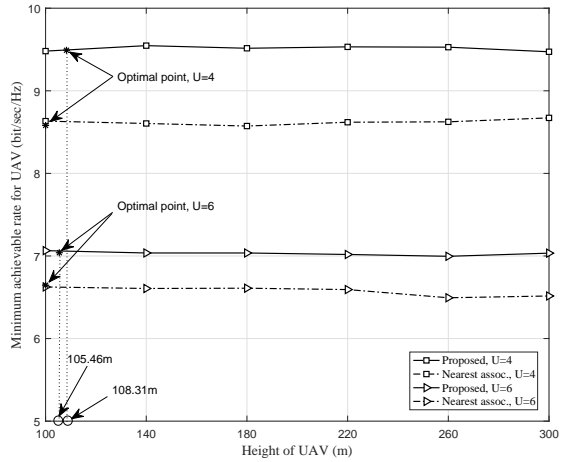,scale=0.44,angle=0}
\end{center}
\caption{Minimum achievable rate for UAV versus flying height of UAV, where $N=14$.}\label{F3}
\end{figure}

With perfect CSI estimation, we mainly demonstrate the optimal UAV flying height for our proposed algorithm and also examine the optimal average UAVs' minimum achievable rate for various number of co-existed G-UEs per cell and number of activated UAVs configurations. The target G-UE's SINR in this subsection is still set to $\gamma = 2$. As shown in Fig.~\ref{F3}, 
the UAVs' minimum achievable rates for various UAV flying heights and our optimized UAV height are almost the same, where for $U=4$, the optimal average UAV flying height is 108.31m in our proposed case and is 100.02m in the nearest association case, and for $U=6$, the optimal average UAV flying height is 105.46m in our proposed case and is 100.02m in the nearest association case. Unlike existing UAV height optimization work, with the proposed beamforming design and LoS UAV channel model (i.e., $h_{u}\geq100\text{m}$), the average UAVs' minimum achievable rate is not sensitive to their flying height. In addition, in comparison to the nearest UAV association case, our proposed UAV association method leads to increased UAV rate performances. 

\begin{figure}[t] \setlength{\belowcaptionskip}{-.5cm}
\begin{center}
\epsfig{figure=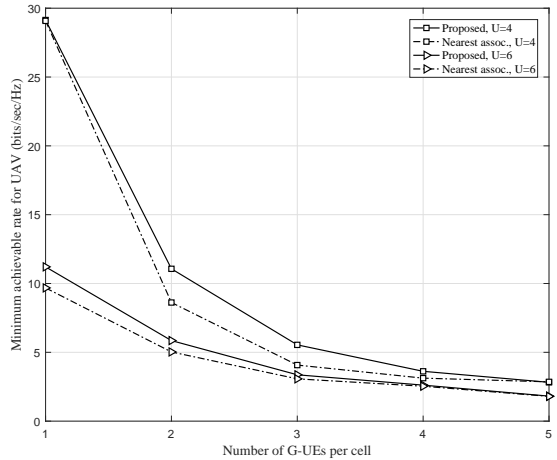,scale=0.44,angle=0}
\end{center}
\caption{Minimum achievable rate for UAV versus number of G-UEs per cell, where $N=8$.}\label{F4}
\end{figure}

In Fig.~\ref{F4}, we compare the UAVs' minimum achievable rate for various number of G-UEs per cell and various number of activated UAVs. 
We can see that the UAVs' minimum achievable rate decreases with increasing the number of G-UEs per cell for $U=4$ and $U=6$. This is because of the increased number of intra-cell and inter-cell interference from G-UEs. In addition, our proposed scheme outperforms the nearest UAV association for various number of G-UEs configurations. The performance gap reduces with increasing the number of G-UEs per cell due to the increased co-channel interference. It is worth noting that, when $U=4$ and $K=1$, the UAVs' minimum achievable rate is the same for both association methods. This is because, with only one associated G-UE per cell, each G-BS will have enough DoFs to handle all interference with the help of our proposed beamforming method, thus no UAV association process is needed to further improve the system rate performances.  

%%%%%%%%%%%%%%%%%%%%%%%
\subsection{Achievable Rate with Imperfect CSI and UAV Mobility}
In this subsection, we consider imperfect CSI and mobile UAVs. The target G-UE's SINR is set to $\gamma = 1$, the number of antennas per G-BS is set to $N=8$, and the number of G-UEs per cell and UAVs are set to $K=4$ and $U=4$, respectively. In the first part, we examine different channel estimation errors and different channel correlation coefficients effects on the UAVs' minimum achievable rate performances. In the second part, taking into account the UAV's mobility, we exploit the stability of our proposed method and the trade-off between UAV's rate and computational complexity for updating the optimized variables. 

\begin{figure}[t] \setlength{\belowcaptionskip}{-.5cm}
\begin{center}
\epsfig{figure=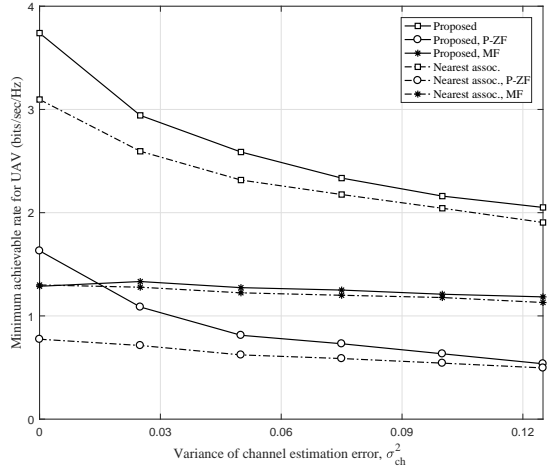,scale=0.44,angle=0}
\end{center}
\caption{Minimum achievable rate for UAV versus variance of channel estimation error, where $\rho=0.6$.}\label{F5}
\end{figure}

\begin{figure}[t] \setlength{\belowcaptionskip}{-.5cm}
\begin{center}
\epsfig{figure=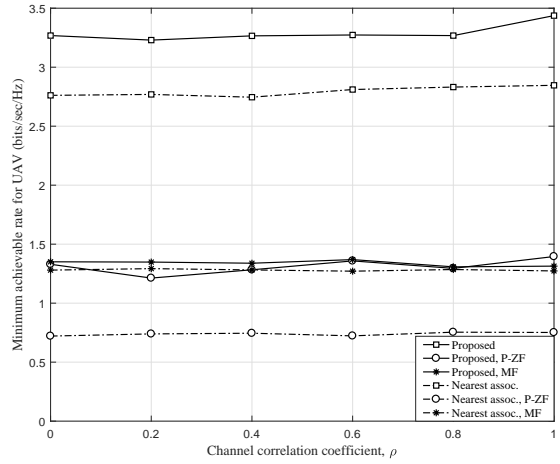,scale=0.44,angle=0}
\end{center}
\caption{Minimum achievable rate for G-UE versus channel correlation coefficient, where $\sigma^{2}_{\mathrm{ch}}=0.0125$.}\label{F6}
\end{figure}

Fig.~\ref{F5} plots the UAVs' minimum achievable rate versus various variance of channel estimation error, i.e., $\sigma^{2}_{\mathrm{ch}}$, where we fix $\rho=0.6$.
We can see that with increasing the channel estimation error, the UAVs' minimum achievable rate decreases. In addition, a large performance gab between our proposed scheme and the nearest UAV association policy can be observed when $\sigma^{2}_{\mathrm{ch}}$ is small. Moreover, unlike the perfect CSI scenario, our proposed algorithm with imperfect CSI shows better performances than the nearest association policy for P-ZF, MF, and our optimized beamforming designs, which showcase the robustness of proposed scheme in term of CSI imperfection. In Fig.~\ref{F6}, we examine the UAVs' minimum rate performance for various channel correlation coefficients, i.e., $\rho$, setup and the fixed $\sigma^{2}_{\mathrm{ch}}=0.0125$.
In this case, by increasing the value of $\rho$, almost the same UAV's rate performance can be observed expect for the case that $\rho=1$, which illustrates that $\rho$ is not quite sensitivity to UAV's rate performance when the channel estimation error $\sigma^2_{\mathrm{ch}}$ is relatively small.

\begin{figure}[t] \setlength{\belowcaptionskip}{-.5cm}
\begin{center}
\epsfig{figure=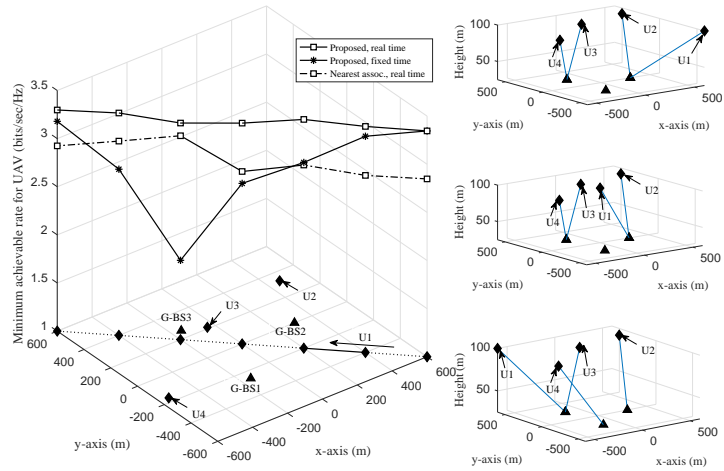,scale=0.46,angle=0}
\end{center}
\caption{Minimum achievable rate for UAV versus UAV's mobility, where $\rho=0.6$, $\sigma^{2}_{\mathrm{ch}}=0.0125$, and $h_{u}=100\text{m},\forall u\in\mathcal{U}$.}\label{F7}
\end{figure}

Considering mobile UAVs, it is almost impossible to keep updating UAV association matrix and beamforming vectors for every time instant due to computational complexity of optimization process. This motivates us to examine the trade-off between the UAVs' minimum achievable rate performance and the frequency for updating the optimization variables. For ease of description, in Fig.~\ref{F7},
we assume there is only one mobile UAV (i.e., U1) is moving in a line from a geometric point (e.g., (600,-600,100)) to a geometric point (e.g., (-600,600,100)). Then, we evenly pick up seven points in the line for carrying out the optimization process for both our proposed algorithm and nearest association method. In the right sub-figures of Fig.~\ref{F7}, we indicate the optimal UAV association patterns for three U1's locations. As shown in these sub-figures, optimal associated UAVs for G-BSs is changing when U1 is moving. In the left sub-figure of Fig.~\ref{F7}, our proposed UAV association optimization (i.e., Proposed, real time) offers quite stable rate performances in comparison to the nearest association method (i.e., Nearest assoc., real time). In addition, to reduce the computational complexity, we also plot the curve named `Proposed, fixed time', in which the optimization process is only conducted at the first point, i.e., (600,-600,100), and the same optimized variables are used to calculate the UAVs' minimum achievable rate in other points. It is shown that, the UAV's rate performance decreases as the UAV is moving, and when the UAV is moving to the point with location (-200,200,100), the worst performance is observed. As the UAV continously moving forward, its minimum achievable rate increases. This is because that, although the out of date optimized variables are used, the symmetric moving pattern of the UAV leads to similar large-scale fading effect. Here, the small-scale fading has already been averaged out. Such observation illustrates the applicability of out-of-date optimized variables for symmetric UAV movement with reduced computational complexity. %, which confirms that the nearest association may not be not a good option, especially when MIMO beamforming is implemented.
%%%%%%%%%%%%%%%%%%%%%%%%%%%%%%%%%%%%%%%%%%%%%%%%%%%%%%%%%%%%%%%%%%%%%%

                         %%% VI. Conclusion %%%

%%%%%%%%%%%%%%%%%%%%%%%%%%%%%%%%%%%%%%%%%%%%%%%%%%%%%%%%%%%%%%%%%%%%%%
\section{Conclusion}
In this paper, we have proposed a joint beamforming, UAV association, and power control framework to maximize the minimum achievable rate for UAV subject to the co-existed G-UEs' target rate being guaranteed. By jointly optimizing the UAV association matrix, MIMO beamforming vectors, and the height of UAVs, the uplink co-channel interference among UAVs and G-UEs can be well eliminated. Taking into account the high mobility of UAVs, we have also revised our proposed method for application in imperfect CSI and mobile UAV scenario. It has been shown that our proposed algorithm can offer remarkable UAVs' minimum achievable rate in comparison to the conventional nearest UAV association policy for both perfect CSI and imperfect CSI scenarios. Moreover, some interesting findings, such as the importance of symmetric UAV path planning, have been observed when we analysed the trade-off between UAVs' minimum rate performance and optimization complexity.

%%%%%%%%%%%%%%%%%%%%%%%%%%%%%%%%%%%%%%%%%%%%%%%%%%%%%%%%%%%%%%%%%%%%%%

                         %%%Acknowledgment%%%

%%%%%%%%%%%%%%%%%%%%%%%%%%%%%%%%%%%%%%%%%%%%%%%%%%%%%%%%%%%%%%%%%%%%%%
%\section*{Acknowledgment}
%The authors would like to thank...% the Editor Prof. and anonymous reviewers for their efficient review process and constructive comments.
%%%%%%%%%%%%%%%%%%%%%%%%%%%%%%%%%%%%%%%%%%%%%%%%%%%%%%%%%%%%%%%%%%%%%%
\section*{Appendix A\\{Find the Big-$M$}}
The value of $M$ must satisfy the following constraints for all $i\in\mathcal{G}$ and for all $u\in\mathcal{U}$, which is
\begin{equation}\label{Appx-A-01}
M\geq t\sum_{k\in\mathcal{K}}\mathcal{D}^{(u)}_{i,k}+t\sum_{g\in\mathcal{G}}\sum_{u^{'}\in\mathcal{U}/u}a_{g,u^{'}}P_{g,u^{'}}\mathcal{F}^{(u)}_{i,u^{'}}+t\mathcal{N}^{(u)}_{i}.
\end{equation}
From \eqref{Appx-A-01} we can see that the value of $M$ depends on $i$, $u$, and $\mathbf{A}$, and without loss of generality, $M$ can be
\begin{equation}\label{Appx-A-02}
M = \underset{\{i\},\{u\},\mathbf{A}}{\max}\left(t\sum_{k\in\mathcal{K}}\mathcal{D}^{(u)}_{i,k}+t\sum_{g\in\mathcal{G}}\sum_{u^{'}\in\mathcal{U}/u}a_{g,u^{'}}P_{g,u^{'}}\mathcal{F}^{(u)}_{i,u^{'}}+t\mathcal{N}^{(u)}_{i}\right),
\end{equation}
According to \eqref{Appx-A-02}, there must exist an associated $i^{*}\in\mathcal{G}$ and $u^{*}\in\mathcal{U}$ such that
\begin{equation}\label{Appx-A-03}
M = t\sum_{k\in\mathcal{K}}\mathcal{D}^{(u^{*})}_{i^{*},k}+t\sum_{g\in\mathcal{G}}\sum_{u^{'}\in\mathcal{U}/u^{*}}a_{g,u^{'}}P_{g,u^{'}}\mathcal{F}^{(u^{*})}_{i^{*},u^{'}}+t\mathcal{N}^{(u^{*})}_{i^{*}}.
\end{equation}
This associated $i^{*}\in\mathcal{G}$ and $u^{*}\in\mathcal{U}$ can be found by comparing $U\times G$ combination values of right-hand side of \eqref{Appx-A-02}. Specifically, we have $U\times G$ combination values for $\sum_{k\in\mathcal{K}}\mathcal{D}^{(u)}_{i,k}$ in \eqref{Appx-A-02} in terms of $i\in\mathcal{G}$ and $u\in\mathcal{U}$. With each combination of $i\in\mathcal{G}$ and $u\in\mathcal{U}$, we can sum up the $U-1$ maximum combination values amount all $G\times(U-1)$ combination values for $P_{g,u^{'}}\mathcal{F}^{(u)}_{i,u^{'}}$ in \eqref{Appx-A-02} in terms of $g\in\mathcal{G}$ and $u'\in\mathcal{U}/u$. Then, by adding $\sum_{k\in\mathcal{K}}\mathcal{D}^{(u)}_{i,k}$ to the corresponding summed $U-1$ maximum combination values for $P_{g,u^{'}}\mathcal{F}^{(u)}_{i,u^{'}}$ in \eqref{Appx-A-02}, we obtain one of $U\times G$ combination values for comparison. Finally, $M$ will be the maximum value among the $U\times G$ calculated combinations.

%%%%%%%%%%%%%%%%%%%%%%%%%%%%%%%%%%%%%%%%%
\bibliography{mybib}
\bibliographystyle{IEEEtran}
\end{document}